\documentclass[twocolumn,twocolappendix]{aastex63}
\graphicspath{{./Figs/}{./png/}}
\usepackage{amsmath}
\usepackage{multirow}

\newcommand{\BB}{\mathbf{B}}
\newcommand{\K}{\,{\rm K}}
\newcommand{\G}{\,{\rm G}}
\newcommand{\uG}{\,\mu{\rm G}}

\newcommand{\g}{\,{\rm g}}
\newcommand{\cm}{\,{\rm cm}}

\newcommand{\cMpch}{\,h^{-1}{\rm cMpc}}

\newcommand{\s}{\,{\rm s}}
\newcommand{\cMpc}{\,{\rm cMpc}}
\newcommand{\ckpc}{\,{\rm ckpc}}
\newcommand{\ckpch}{\,h^{-1}{\rm ckpc}}
\newcommand{\ckpchI}{\,h\,{\rm ckpc^{-1}}}
\newcommand{\cMpchI}{\,h\,{\rm cMpc^{-1}}}

\newcommand{\Gyr}{\,{\rm Gyr}}

\shorttitle{Primordial Magnetic fields in Galaxy Clusters}
\shortauthors{Mtchedlidze et al.}
\graphicspath{{./}{figures/}}

\begin{document}

\title{Inflationary and phase-transitional primordial magnetic fields in galaxy clusters}

\author[0000-0001-9786-8882]{Salome Mtchedlidze}
\affiliation{School of Natural Sciences and Medicine, Ilia State University, 3-5 Cholokashvili St., 0194 Tbilisi, Georgia}
\affiliation{
Institute for Astrophysics and Geophysics,
Georg-August-Universit\"at G\"ottingen, Friedrich-Hund-Platz 1, D-37077 G\"ottingen, Germany}
\affiliation{E. Kharadze Georgian National Astrophysical Observatory, 47-57 Kostava St., 0179 Tbilisi,Georgia}
\email{salome.mtchedlidze.1@iliauni.edu.ge}

\author[0000-0001-7058-8418]{Paola Dom\'inguez-Fern\'andez}
\affiliation{Dipartimento di Fisica e Astronomia, Universit\'{a} di Bologna, Via Gobetti 92/3, 40121, Bologna, Italy}
\affiliation{Hamburger Sternwarte, Universit\"at Hamburg, Gojenbergsweg 112, 21029 Hamburg, Germany}

\author[0000-0003-0728-2533]{Xiaolong Du}
\affiliation{Carnegie Observatories, 813 Santa Barbara Street, Pasadena, CA 91101, USA}

\author[0000-0001-5233-8087]{Wolfram Schmidt}
\affiliation{Hamburger Sternwarte, Universit\"at Hamburg, Gojenbergsweg 112, 21029 Hamburg, Germany}

\author[0000-0002-7304-021X]{Axel Brandenburg}
\affiliation{Nordita, KTH Royal Institute of Technology and Stockholm University, Hannes Alfv\'ens v\"ag 12, SE-10691 Stockholm, Sweden}
\affiliation{The Oskar Klein Centre, Department of Astronomy, Stockholm University, AlbaNova, SE-10691 Stockholm, Sweden}
\affiliation{McWilliams Center for Cosmology and Department of Physics, Carnegie Mellon University, 5000 Forbes Ave, Pittsburgh, PA 15213, USA}
\affiliation{School of Natural Sciences and Medicine, Ilia State University, 3-5 Cholokashvili St., 0194 Tbilisi, Georgia}

\author[0000-0002-3063-4325]{Jens Niemeyer}
\affiliation{
Institute for Astrophysics and Geophysics,
Georg-August-Universit\"at G\"ottingen, Friedrich-Hund-Platz 1, D-37077 G\"ottingen, Germany}

\author[0000-0003-0217-9852]{Tina Kahniashvili}
\affiliation{McWilliams Center for Cosmology and Department of Physics, Carnegie Mellon University, 5000 Forbes Ave, Pittsburgh, PA 15213, USA}
\affiliation{School of Natural Sciences and Medicine, Ilia State University, 3-5 Cholokashvili St., 0194 Tbilisi, Georgia}
\affiliation{E. Kharadze Georgian National Astrophysical Observatory, 0179, 47-57 Kostava St., Tbilisi,Georgia}

\begin{abstract}

Primordial magnetic fields (PMFs) are possible candidates for explaining
the observed magnetic fields in galaxy clusters.
Two competing scenarios of primordial magnetogenesis have been discussed
in the literature: inflationary and phase-transitional.
We study the amplification of both large- and small-scale
correlated magnetic fields, corresponding to inflation- and
phase transition-generated PMFs, in a massive galaxy cluster.
We employ high-resolution
magnetohydrodynamic cosmological zoom-in simulations to
resolve the turbulent motions in the intracluster medium.
We find that the turbulent amplification
is more efficient for the large-scale inflationary models, while
the phase transition-generated seed fields show moderate growth. 
The differences between the models are imprinted on the spectral
characteristics of the field (such as the amplitude and the shape of the magnetic
power spectrum) and therefore on the final correlation length.
We find a one order of magnitude difference between the final strengths
of the inflation- and phase transition-generated magnetic fields,
and a factor of $1.5$ difference between their final coherence scales.
Thus, the final configuration of the magnetic field retains information
about the PMF generation scenarios.
Our findings have implications for future extragalactic Faraday rotation
surveys with the possibility of distinguishing between different
magnetogenesis scenarios.

\end{abstract}


\section{Introduction} \label{sec:intro}

Galaxy clusters, the largest virialized structures of the universe,
reveal the existence of large-scale correlated magnetic fields in the
dilute plasma between galaxies that is known as the intracluster medium (ICM).
Studies of Faraday rotation measures, as well as diffuse radio emissions
in the form of radio halos and radio relics, have probed the strength
and morphology of the ICM magnetic field \citep[see, e.g.,][for
reviews]{GovFer2004,Brueggenetal2012,Weerenetal2019}.
These different observational methods infer a field strength of the order
of microGauss and coherence scales reaching a few tens of kiloparsecs in galaxy
clusters \citep[see, e.g.,][]{GovFer2004,Weerenetal2019}.

Despite their ubiquity, the origins of cluster magnetic fields remain elusive.
A commonly accepted hypothesis is that weak seed magnetic fields,
generated from an initially zero magnetic field \citep{Rees1987},
are amplified during structure formation via the combined effects of
adiabatic compression and a small-scale dynamo \citep[see, e.g., the recent review by][]{Donnertetal2018}.
It is debatable whether these seed magnetic fields are produced
in the early universe by the primordial magnetogenesis or whether they are produced at
a later epoch, during structure formation, by astrophysical
mechanisms \citep[e.g., the Biermann battery mechanism and the Weibel
instability---][]{Biermann1950, Lazaretal2009}.
In the primordial scenario, magnetic fields originating in the early
universe, i.e., primordial magnetic Fields (PMFs), have volume-filling
fractions that are close to unity, making them good candidates for explaining the
magnetization of cosmic voids.
This scenario is favored by observations of blazar spectra that rule out
the possibility of a zero magnetic field in the intergalactic medium \citep[IGM; see][and references therein]{Fermi_Lat2018}.
On the other hand, the astrophysical scenario requires efficient transport
mechanisms of magnetic energy toward larger scales, to explain the
possible magnetization of cosmic voids.
Galactic winds \citep[][]{Kronbergetal1999,Bertoneetal2006} as well as
jets and lobes from radio galaxies \citep{DalyLoeb1990} have been proposed
as such efficient transport processes.
However, the significance of the volume-filling factor of such a locally
generated and transported magnetic field remains unclear \citep[see,
e.g.,][]{Dolagetal2011,Bondarenkoetal2021}.

PMFs could result from different magnetogenesis scenarios.
Their post-recombination magnetic structure 
and the field coherence scale
depend on: (1) the details of the particular magnetogenesis model; and (2)
evolutionary trends in the pre-recombination universe.
A primordial seed field could be generated during
inflation or phase transitions \citep[see][for recent
reviews]{Subramanian2016,Vachaspati2020}.
In inflationary magnetogenesis, the coherence scale of the
quantum-mechanically produced seed magnetic field 
can be stretched on superhorizon scales.
However, the conformal invariance of the electromagnetic action must be broken
\citep[see, e.g.,][]{Dolgov1993} in order to achieve sufficiently strong seed fields
for their subsequent growth at later epochs.
This is usually ensured by the coupling of the electromagnetic action with
scalar fields, such as the inflaton \citep{TurnerWidrow1988,Ratra1992},
or by nonminimal coupling with the scalar-tensor gravity, as has
been proposed by \cite{Mukohyama2016}.
Contrary to the inflationary scenario, the coherence scale of the phase transition-generated field is 
limited by the Hubble horizon scale, and it is a sizeable fraction of the Hubble scale.
The electroweak (EW) or quantum-chromodynamical (QCD) phase transitions
produce seed fields through nonequilibrium processes, e.g., during
the collision \citep[][]{AhonenEnqvist1998,Copelandetal2000} and nucleation
\citep[][]{ChengOlinto1994,Sigletal1997} of bubbles of different phases
\citep[see][for a review]{Kandusetal2011}.
Both inflation- and phase transition-generated seed fields are assumed
to have a stochastic distribution.
In addition, in the inflationary scenario, the constant, spatially
uniform magnetic field is predicted by the Mukohyama model
\citep{Mukohyama2016,Mandaletal2020}.

The evolution of PMFs proceeds as ``freely decaying turbulence''
in the radiation-dominated epoch \citep[see, e.g.,][]{Brandenburgetal_1996,
Christenssonetal2001, BanerjeeJedamzik2004, Kahniashvilietal2016,
Brandenburgetal2018}.
The correlation length of the small-scale (phase-transitional) as well as the
large-scale (inflationary) correlated primordial field increases, although
much more efficiently in the former case, due to an inverse cascade
\citep{Kahniashvilietal2010,Brandenburgetal2018}.
The field is expected to freeze and retain its characteristic spectral
profile, from the moment of recombination until reionization.
The formation of massive structures, such as galaxy clusters,
and the subsequent amplification of PMFs on the corresponding scales
\citep[see][for a review]{Donnertetal2018}, can be studied with
cosmological magnetohydrodynamic (MHD) simulations \citep[see,
e.g.,][]{Dolagetal1999a, DubTey2008, Xuetal2009, Vazzaetal2014,
Marinnacietal2018}.
In the present paper, we study the amplification of the two types of
PMFs in a massive galaxy cluster with the \texttt{Enzo} code
\citep{Bryanetal2014}, using the adaptive mesh refinement (AMR) technique.
For the first time, we compare the evolution of small- and
large-scale correlated PMFs, consistent with different inflationary and
phase-transitional primordial magnetogenesis scenarios.

The structure of this paper is as follows.
In Section~\ref{sec:PhysMod} we discuss our physical model and our motivation for studying different PMFs.
In Section~\ref{sec:simulations}, we describe our numerical setup, the initial conditions, and the details of the simulated galaxy cluster.
In Section~\ref{sec:Results}, we present our results.
Finally, we discuss possible numerical caveats in Section~\ref{sec:NumAsp}
and we summarize our main results in Section~\ref{sec:Summ}.

\section{Physical model}
\label{sec:PhysMod}

In this section, we describe the spectral characteristics of the inflation-
and phase transition-generated PMFs that are used as initial conditions in our
simulations.
Regardless of the magnetogenesis scenario, it is generally expected
that after generation, the primordial seed magnetic field is
frozen in, and the amplitude of the (physical) magnetic
field decreases with the expansion of the universe, $B_{\text{phys}}
\propto 1/a^2$; equivalently, magnetic field lines are adiabatically
stretched, with a $\propto \rho_{\text{phys}}^{2/3}$ scaling, where
$\rho_{\text{phys}}$ is the gas density.
This treatment of the magnetic field may be justified in a highly
conducting fluid, such as the hot plasma in the early universe, if there
is no turbulence.
However, the concept of simple adiabatic dilution has to be abandoned, when the effects of turbulence can become important.
Turbulent magnetic fields could be generated
either at the end of inflation through
the inflaton decay to standard model fields, or during
phase transitions, through collisions between the expanding bubbles of the new phase
\citep[see][for reviews]{Subramanian2016,Vachaspati2020},
and they could
alter the evolution of different primordial seed magnetic fields.
Random magnetic fields could also be generated when the phase transition
only involves a smooth crossover to the new phase, without bubbles.
In the following, we will discuss how the statistical properties
of the inflation- and phase transition-generated magnetic fields are
modified when taking into account their turbulent (pre-recombination)
evolution after their generation.

In the statistical framework, the description of PMFs relies on the
definition of the magnetic energy power spectra and their characteristic length scales \citep[also see the discussion in Section~3 of] [hereafter, Paper~I]{Mtchedlidzeetal_2022}.
The magnetic energy power spectrum $E_{B}(k)$ is often conveniently
split into its large-scale, $E_{B}^{\text{LS}}$ and small-scale,
$E_{B}^{\text{SS}}$ parts.
The transition from the large-scale to the small-scale spectra occurs
at the scale corresponding to the wavenumber $k_{\rm peak}$. In the case of phase transition-generated PMF, this scale corresponds
to the phase transition bubble size and cannot exceed the Hubble
horizon size at the moment of field generation
\citep[see, e.g.,][]{Kahniashvilietal2010}.
After their generation, the decaying turbulence leads to a magnetic
energy spectrum, which can either be $E_{B}^{\text{LS}}
\propto k^2$, commonly known as the ``\textit{Saffman spectrum}'' \citep{Hogan1983},
or $E_{B}^{\text{LS}} \propto k^4$, known as the ``\textit{Batchelor
spectrum}'' \citep{Davidson2004}.
As \cite{Kahniashvilietal2010} have shown, the realization of
these spectra depends on the driving nature of the turbulent magnetic field.
If the turbulence is driven through kinetic energy injection, magnetic
field develops a spectrum close to $ E_{B}^{\text{LS}} \propto k^4$
(Batchelor) spectrum; if the initial driver is a magnetic field,
then the spectrum is shallower than $E_{B}^{\text{LS}} \propto k^3$.
In addition, the recent work of \cite{BZS22} has shown that in the former
scenario (weak magnetic fields), the Batchelor and Saffman spectra result
from small-scale dynamo action, in its kinematic and saturated states,
respectively.
The Batchelor spectrum is also expected from the causality
condition being combined with the divergence-free field condition
\citep{Durrer+Caprini2003}.
Finally, on smaller scales, a turbulent magnetic cascade with
$E_{B}^{\text{SS}} \propto k^{-5/3}$ is expected for both the
Saffman and Batchelor spectra.

The inflationary scenario, in turn, predicts a magnetic energy spectrum
that can be nearly scale-invariant at the moment of generation,
i.e., $E_{B}(k) \propto k^{-1}$.
This scaling is further modified, due to turbulent decay during the
pre-recombination epoch, and results in a Kolmogorov $k^{-5/3}$ spectrum
by the end of recombination \citep{Kahniashvilietal2017}.
It should be noted that a transition to the $k^4$ spectrum (IR cutoff) could also be a possible outcome of inflation, although on much
larger scales than the characteristic scale for the phase-transitional
scenarios, that is, $k_{\rm peak}^{\rm infl} \ll k_{\rm peak}^{\rm PT}$
\citep[see, e.g.,][]{Brandenburgetal2018}.
In this case, \cite{Brandenburgetal2018} found that for a
certain wavenumber range (close to the peak of the spectrum,
$k>k_{\rm peak}^{\rm infl}$), the power spectrum will still be characterized by the
scale-invariant spectrum.

In the present work, we explore two phase transition-generated
PMFs that are characterized by a Saffman spectrum and a Batchelor spectrum, respectively, and two
inflationary-generated PMFs that are characterized by a turbulent spectrum and
by a Dirac delta function (in Fourier space, corresponding to a
uniform magnetic field), respectively.
The latter model serves as a comparison to our simulations
with other cosmological simulations, where a uniform seed magnetic
field is commonly assumed as an initial condition \citep[see,
e.g.,][]{Dolagetal1999b,Marinnacietal2015,Vazzaetal2018}.
Nevertheless, the physical generation of a uniform seed magnetic field
in the early universe has been predicted to be plausible under specific
conditions by \cite{Mukohyama2016}.
Hereafter, we refer to this model as the Mukohyama model and we also refer
the reader to \cite{Mandaletal2020}, and the references therein,
for more details.

We adopt these models as our initial magnetic conditions, despite our
relatively low initial resolution of $312.5\,\ckpch$, where the `c'
is commonly used to emphasize comoving units.
We note that this initial resolution may not be enough to resolve
the magnetic field coherence scales that are expected from theory or the small scales that are dominated by the turbulent spectra.
For example, in the phase-transitional scenario, an optimistic assumption
of the largest magnetic eddy size would lead to magnetic field coherence
scales of the order of $10$ kpc (comoving) at the end of recombination
\citep[see, e.g., the constraint plot, Figure~7, in][]{RoperPoletal2022}.
However, \cite{Kahniashvilietal2022} have recently proposed that 
QCD phase transition-generated PMFs could
even reach $\sim 300 \ckpc$ coherence scales (if the field is fully
helical), by accounting for the decaying nature of turbulent sources
between the time of generation and big bang nucleosynthesis.
The hypothesis behind this finding is that the magnetic correlation length
can be larger if one applies the Big Bang nucleosynthesis (BBN) limits not
to the time of generation of the seed field, but to the later time of the BBN.
While the predicted magnetic field coherence scale may vary from theory
to theory, we emphasize that our initial resolution prevents us from having
a one-to-one match with any of the various theoretical expectations.
It is therefore important to stress that, similar to Paper~I, our
initial stochastic spectra are only intended to emulate the shapes that
are theoretically expected.

\section{Simulations} \label{sec:simulations}

We simulate the formation of a galaxy cluster with the cosmological
Eulerian MHD code \texttt{Enzo} \citep{Bryanetal2014}.
We assume a Lambda cold dark matter ($\Lambda$CDM) cosmology ($h=0.674$,
$\Omega_m=0.315$, $\Omega_b=0.0493$, $\Omega_{\Lambda}=0.685$, and
$\sigma_8=0.807$, as in the \citealt{Planck2018}).
As in our previous work (Paper~I)\footnote{We refer the reader to this
paper for a detailed description of the adopted temporal and spatial
reconstruction schemes.}, we use the Dedner formulation of the MHD
equations, to obey the divergence-free condition of the magnetic field
\citep[][]{Dedneretal2002}.
In the present paper, we additionally employ AMR to reach a higher resolution within our simulated galaxy cluster
\citep{2019JOSS....4.1636B}.

We follow two steps to solve the galaxy cluster: (1) a global AMR simulation, where we identify a list of fairly resolved haloes
and (2) a local AMR or ``zoom-in'' simulation, where we apply
several levels of AMR in a selected region in which the cluster forms.
In both setups, the refinement is triggered according to the baryon, $f_{\text{b}}$,
and dark matter (DM), $f_{\text{DM}}$, overdensity thresholds.
These parameters ensure refinement when the gas (DM) mass in a
cell reaches a factor of $f_{\text{b}}$ ($f_{\text{DM}}$) times
the mean baryonic (DM) mass expected in a cell at the root grid level
\citep{Bryanetal2014}.
In this study, we use a nominal refinement factor of 2 between the parent
grid and its subgrid, which is the commonly used value for cosmological
simulations \citep[see][for more details]{Bryanetal2014}.
In the global AMR simulations, we set $f_{\text{b}} = f_{\text{DM}} = 4$
and we use 4 levels of refinement that are activated in the whole $(80\,\cMpch)^3$
simulation box.
We use a root grid of $256^3$ cells and $256^3$ DM
particles each of mass $m_{\text{DM}} = 3.34\times 10^{9} M_{\odot}$.
The initial and final spatial resolutions are $312.5\,\ckpch$ and final
$19.5\,\ckpch$, respectively.
Based on this simulation, we produce a halo catalog using the \textit{yt}
halo finder \citep{SkoryTurk2011}.
The halo finder identifies groups of linked DM particles, based on the
\cite{EisHut1998} algorithm. The galaxy cluster selected for the present
work is among the most massive clusters from our halo catalog (see
Section~\ref{subsec:selCluster} for a detail description of the cluster).
Next, we resimulate the selected galaxy cluster in the $(80\,\cMpch)^3$
simulation box, by centering our simulation box where the galaxy cluster
forms.
We select a volume of $(20\,\cMpch)^3$ and use seven levels of refinement.
In this case, the refinement is triggered on the $f_{\text{b}}=0.1$ and
$f_{\text{DM}}=4$ refinement factors, giving us a final maximum spatial
resolution of $2.44\,\ckpch$.

\begin{deluxetable*}{c| c c c c c c c}
\label{tab:Tab1}
\tablecaption{
Initial conditions for the magnetic field. 
The correlation length and the mean value of the smoothed
(on a $1 \cMpch$ scale) magnetic field are denoted
by $\lambda_{B}$ and $B_{1 \rm Mpc}$, respectively, while
$\langle B_{0}^2 \rangle$ and $\langle B_{0} \rangle$ are the
means of the initial magnetic field energy and the initial
magnetic field strength, respectively\tablenotemark{a}.
}
\tablehead{
\colhead{Scenario} & \colhead{Model} &  \colhead{Simulation ID}& \colhead{$\langle B_{0}^2 \rangle$}  & \colhead{$\langle B_{0} \rangle$} &  \colhead{$B_{1 \rm Mpc}$}&\colhead{$\lambda_{B}$} \\
\colhead{} &\colhead{}&\colhead{~}  & \colhead{ [(nG)$^2$]}  &\colhead{[nG]} &\colhead{[nG]} & \colhead{[$h^{-1}\cMpc$]} 
}
\startdata
\multirow{2}{*}{Inflationary} & \multirow{1}{*}{(i) Uniform}  &  u  & 0.99 & 0.99 &---&---\\
            & \multirow{1}{*}{(ii) Scale-invariant}  & km1 & 0.99 & 0.92 & 0.92 & 33.04\\
\hline
\multirow{2}{*}{ Phase-transitional} & \multirow{1}{*}{(iii) Saffman} &  k2  & 0.99 & 0.92 & 0.92 & 1.07  \\
            & \multirow{1}{*}{(iv) Batchelor}  & k4  & 0.99 & 0.92 & 0.92 & 0.85\\
\enddata
\tablenotetext{a}{We use comoving quantities everywhere unless stated otherwise.}
\end{deluxetable*}

The selection of the overdensity factors, $f_i$ (where ``i'' indicates
baryons or DM), is important, and it depends on the problem being addressed.
In this work, the grid refinement thresholds are chosen in order to
solve the turbulent motions in the ICM that are
crucial for the seed magnetic field amplification.
Mergers and accretion events that are driven by gravitational dynamics are the
main agents of turbulence in the ICM.
Therefore, low overdensity thresholds for both gas and DM ensure
resolving low-mass gas substructures and DM halos \citep[as discussed in][]{Osheaetal2005}, and, thus, the maintenance of turbulence in the ICM
\citep[][]{IapNiem_2008}.
Note that lower refinement factors significantly increase the number of
refined grids, so one has to compromise between the final resolution and
the computational cost.
For this purpose, we use a higher value of $f_{\text{DM}}$ compared
to the $f_{\text{b}}$ factor.
This selection closely follows \cite{Vazzaetal2018}, where
the authors have proven that the impact of an increased DM
resolution on the final magnetic field distribution is only minor
\citep[see Figure~17 of][]{Vazzaetal2018}.
Indeed, we will show in Section~\ref{subsec:selCluster} that the chosen
refinement thresholds result in large turbulence-filling factors in our
simulated ICM.

Finally, our simulations do not include gas cooling, chemical evolution,
star formation, or feedback from active galactic nuclei.
As in Paper I, we focus solely on the magnetic field amplification that is due
to structure formation and turbulent flows in the ICM.

\subsection{Initial conditions}
\label{subsec:Inits}

We study four different realizations of the simulated galaxy cluster.
Our simulations differ in the initial magnetic field configurations.
We assume only nonhelical magnetic fields at the initial redshift $z=50$.
Similar to Paper~I, we choose to normalize our initial
magnetic conditions, so that they have the same total magnetic energy (see
Table~\ref{tab:Tab1}). The four models are:

\begin{enumerate}
    \item[1.] 
    Uniform (spatially homogeneous) field: a seed magnetic field with a constant strength across the whole
    computational domain, and directed along the diagonal.
    This case corresponds to a particular inflationary magnetogenesis scenario---namely, the Mukohyama model \citep[][]{Mukohyama2016}.
    \item[2.]
    Scale-invariant field: this is a setup for a stochastic,
    statistically homogeneous PMF, corresponding to an
    inflationary scenario.\footnote{Note that we call this model
    ``scale-invariant," even though it has a turbulent spectrum
    with a $k^{-5/3}$ scaling.
    This is because of the presence of turbulence, which
    quickly changes a $k^{-1}$ spectrum quickly to a $k^{-5/3}$ spectrum, which
    is then the expected outcome after recombination
    \citep[see][for
    details]{Kahniashvilietal2017,Brandenburgetal2018}.
    }
    \item[3.]
    Saffman model: a stochastic, phase transition-generated PMF, which
    has a Saffman spectrum, i.e., with a power-law index of 2.
    \item[4]
    Batchelor model: the same stochastic setup as in (3), but with a Batchelor spectrum, i.e., with a power-law index of 4.
\end{enumerate}

\begin{figure}[t]
    \centering
    \includegraphics[width=9.6cm]{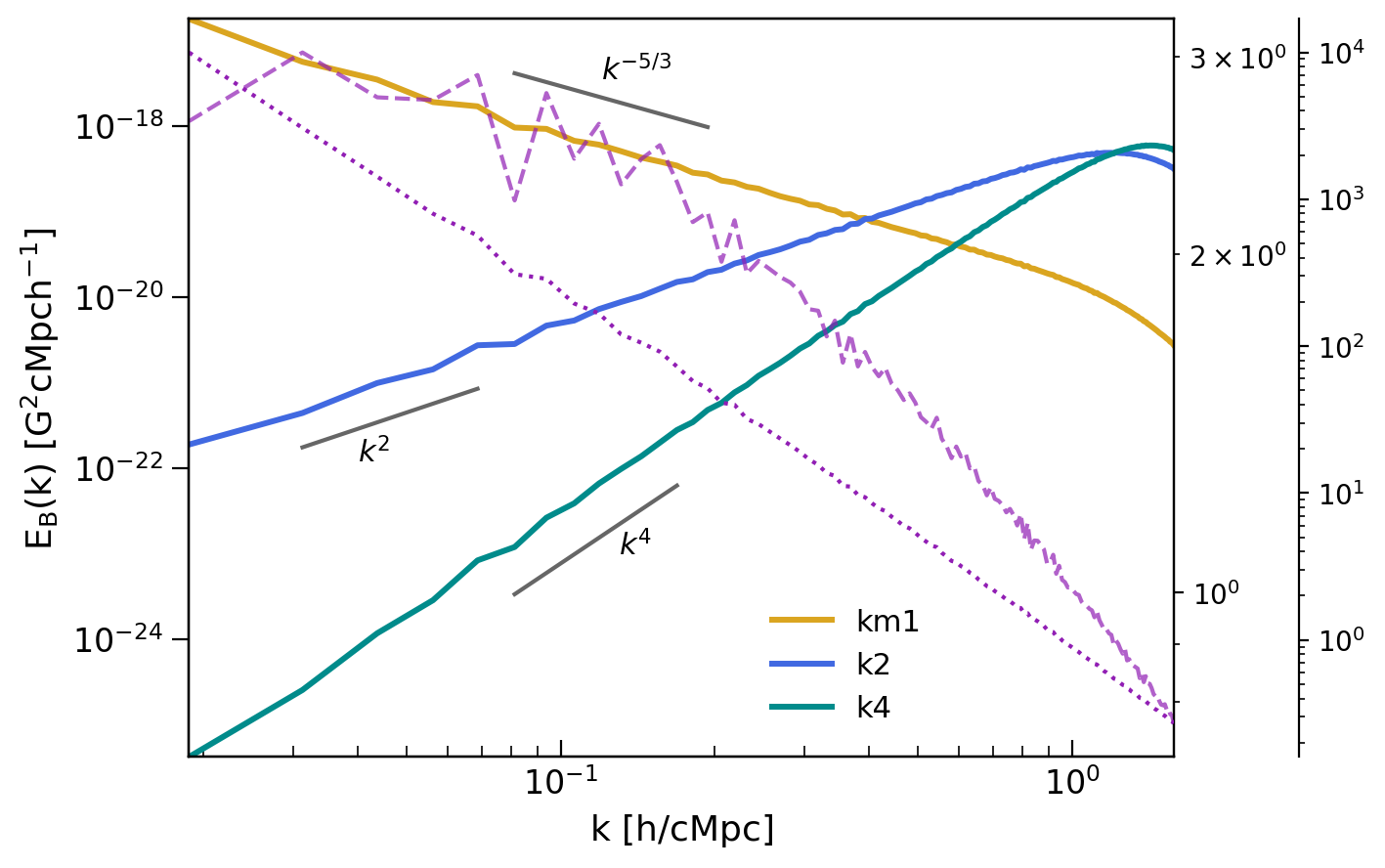}
   \caption{The initial magnetic power spectra for the stochastic setups, with the velocity (dotted purple line) and density (dashed purple line) spectra being shown for the run with the uniform model. The main and second secondary axes shown on the right correspond to the density and velocity spectra, in $(10^{63} \g /\cm^3)^2 \cMpch$ and $10^{8}\cm^2/\s^2 \cMpch$ units, respectively. The initial power spectra of the baryon and DM perturbations are nearly indistinguishable at the scales resolved by our resolution. The only difference between these two spectra is in their amplitudes.}
    \label{fig:B-PS_inits}
\end{figure}

The initial conditions (2)--(4) were produced with the
\textsc{Pencil Code} \citep{JOSS}.
The initial magnetic power spectra for these stochastic setups are shown
in Figure~\ref{fig:B-PS_inits}.
We follow the same method as in Paper~I to generate our initial conditions.
This initial simulation allows us to evolve an initially Gaussian random
field, with the desired spectral properties, until the phase of the magnetic
field in Fourier space become correlated, and their distribution is no
longer one of white noise.
This is then used as the actual initial condition for the \texttt{Enzo} simulations.
The reader may refer to Appendix~A of Paper~I for further details
concerning the generation and normalization of the initial magnetic
conditions (2)--(4).

\begin{figure*}[htbp]
    \centering
    \includegraphics[width=17.7cm]{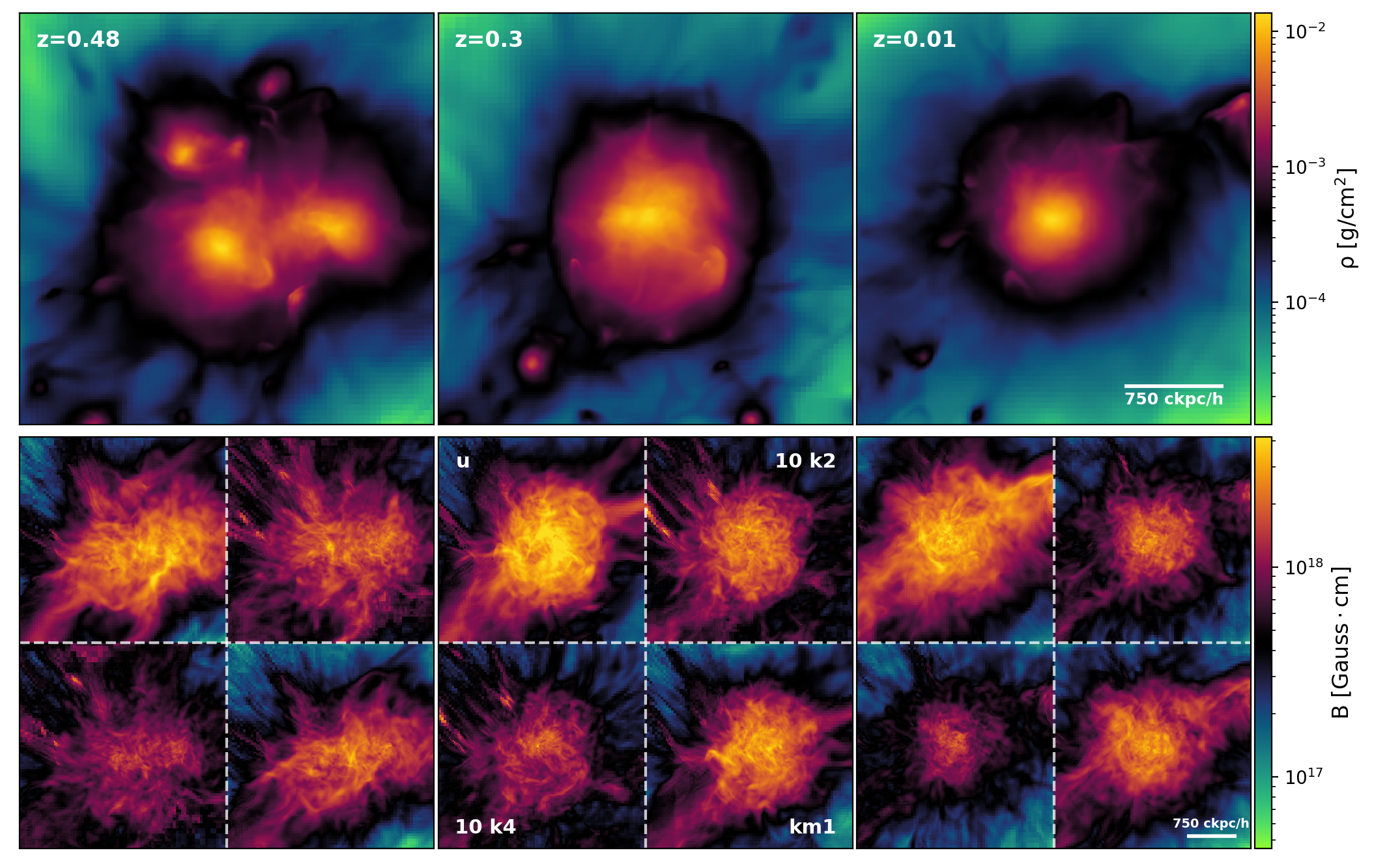}
    \caption{Projected maps of the gas density (top panel) and the magnetic
    field from a $(3\cMpch)^3$ box for different seeding scenarios
    (bottom panel), at different stages of the cluster evolution.
    The left, middle, and right panels show the projected fields at the merging
    ($z=0.48$), post-merger ($z=0.3$) and relaxing ($z=0.01$) states,
    respectively.
    The magnetic field projections for the Batchelor and Saffman models are
    normalized by a factor of 10.
    } 
    \label{fig:projections}
\end{figure*}
We use an initial matter power spectrum, resulting from a primordial,
scale-invariant spectrum, by taking into account the evolution
of post-inflationary linear perturbations, i.e., we use the transfer
function of \cite{EisHu1998}.
It should be noted that the adopted matter power spectrum neglects
any contribution from PMFs.
PMFs are expected to affect the clustering of matter on intermediate and
small scales, i.e., smaller than galaxy cluster scales \citep[][and see
also discussion in Paper~I]{SethiSub2005, Yamazakietal2006,
FedeliMoscardini2012, Kahniashvilietal2013, Sanatietal2020}.
Therefore, we do not expect that the presence of PMF-induced density
perturbations in the early universe to have a significant impact on
our results.

\subsection{Selected cluster}
\label{subsec:selCluster}

The selected cluster from our two-step simulations can be seen in
Figure~\ref{fig:projections}.
The total mass of our cluster, $2.39\cdot 10^{14} M_{\odot}$, is
comparable to the masses of some observed galaxy clusters, such as A3527 \citep[see, e.g.,][]{deGasperinetal_2017} or the recently studied
Ant cluster \citep[][]{Botteonetal2021}.
We summarize the most important parameters of our simulated cluster
in Table~\ref{tab:Tab2}.

\begin{deluxetable}{c c c }
\label{tab:Tab2}
\tablecaption{Characteristics: mass and energy ratio 
$E_{\text{kin}}/E_{\text{tot}}$
of the cluster at $z=0$, where $E_{\text{tot}}= E_{\text{kin}} + E_{\text{th}}$.
}
\tablehead{
\colhead{Radius} & \colhead{Mass} &  \colhead{$E_{\text{kin}}/E_{\text{tot}}$} \\
\colhead{$[\cMpch]$} &\colhead{[$10^{14} M_{\odot}$]} &\colhead{}
 }
\startdata
$R_{500}= 0.50$       &  $1.14$  & $0.15$  \\
$R_{100}= 1.01$       &  $1.86$  & $0.16$  \\
$R_{\text{vir}}=1.54$ &  $2.39$  & $0.16$  \\
\enddata
\end{deluxetable}
%

\begin{figure}[t]
    \centering
    \includegraphics[width=9.2cm]{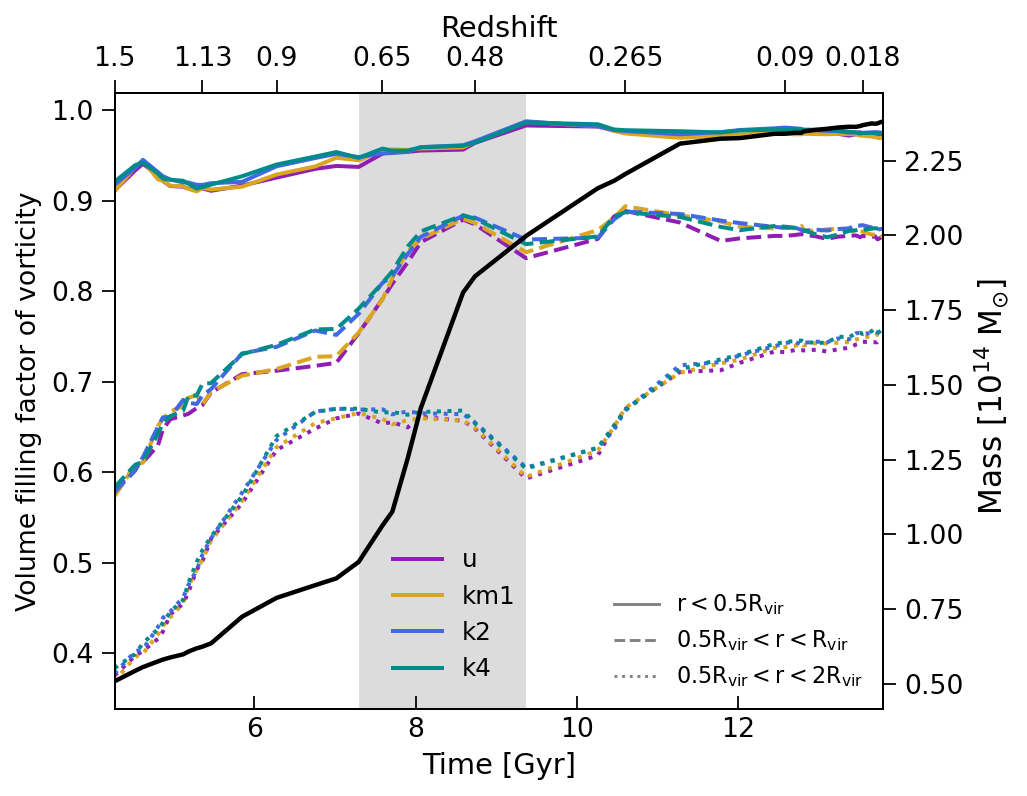}
    \caption{Time evolution of the total virial ($r=R_{\text{vir}}$) mass
    (black solid line) and vorticity volume-filling factor for the cluster
    core (solid lines) and the outskirts enclosing a spherical
    shells in: $0.5 \, R_{\text{vir}} < r <  R_{\text{vir}}$ (dashed lines)
    and  $0.5 \, R_{\text{vir}} < r < 2 \, R_{\text{vir}}$ (dotted lines).
    }
    \label{fig:masses_energy}
\end{figure}

The formation history of a galaxy cluster fully determines the
amount of amplification of the seed magnetic field. 
Our selected cluster undergoes a series of mergers, and its evolution can
be characterized by three phases: (1) at the early stage of formation,
$z\lesssim 0.7$, it continuously grows, by several accreting minor merger
events; (2) in the redshift range 0.7 -- 0.3, a major merger takes place, with a mass ratio of $1.2$ between the main and secondary clusters
(within $R_{500}$ radius), and (3) at late redshifts, i.e., $z<0.3$,
it enters into a relaxing state.
In Figure~\ref{fig:masses_energy}, we show the mass accretion history of
the cluster in the redshift range $1.5>z>0$.
The mass of the cluster is computed within $R_\text{vir}$, and we show
its evolution for the uniform model.
We indicate the major merger phase with the shaded gray area ($\sim 2\Gyr$
timescale) in Figure~\ref{fig:masses_energy}.
During this phase, we observe a steep growth of the total mass, which increases by a factor of $\sim 2$.

Mergers of clusters play a key role in shaping the properties of the
ICM, by injecting turbulence.
To characterize the turbulence in our simulated galaxy cluster,
we follow the recipe proposed by \cite{Iapichinoetal2017}.
In that work, the authors used the vorticity modulus as an indicator
of the velocity fluctuations and its volume-filling factor $f_{\omega}$
as a proxy for the turbulent states of galaxy clusters.
In detail, the procedure consists of flagging a cell as ``turbulent''
if it satisfies the criterion \citep[see][and references
therein]{Kangetal2007,Iapichinoetal2017}
\begin{equation}
\label{eq:omega}
\omega_i > N/t_{\text{age}},
\end{equation}
where $\omega_i$ is the vorticity in the $i$th cell, $t_{\text{age}}$
is the age of the universe at redshift $z$, and $N$ is the number of
eddy turnovers, respectively.
Following \cite{Iapichinoetal2017}, we set $N=10$.
Finally, the volume-filling factor $f_{\omega}$ is the volume fraction
satisfying Equation~(\ref{eq:omega}).
The authors find that $f_{\omega}$ is substantial, both in the core and
at the outskirts of their simulated galaxy cluster, reaching $f_{\omega}>90\%$
and $f_{\omega}>60\%$, respectively.
In the bottom panel of Figure~\ref{fig:masses_energy}, we show the
evolution of the volume-filling factors computed for the core and
outskirt regions of our simulated galaxy cluster.
The volume-filling factors are also shown to be substantial, with
percentages larger than $90\%$ in the core region and $60\%$ in the
outskirts.
We note that we obtain similar results to \cite{Iapichinoetal2017},
even though our numerical setups differ.
For example, their simulations use 8 AMR levels, triggered by spatial
derivatives of the velocity field, to reach a final maximal resolution
of $7.8 \cMpch$.
Additionally, they make use of a subgrid-scale model, which is based on the \cite{Germano92} formalism, to account for unresolved turbulent motions in the ICM;
see also \cite{Schmidtetal2006}.
Thus, our volume-filling factors, along with high final resolution of
$2.44 \ckpch$, show that our numerical setup is adequate for
capturing turbulent motions in the simulated galaxy cluster.

\section{Results}
\label{sec:Results}

\subsection{General properties}
\label{sec:general}

We start our analysis by giving a qualitative view of the density
and magnetic field distribution in the simulated galaxy cluster.
In Figure~\ref{fig:projections}, we show the projected density
and corresponding magnetic field distribution for different seeding
scenarios.
The projections are extracted from a $(3 \cMpch)^3$ simulation box, for three different epochs: the merging ($z=0.48$), post-merging ($z=0.3$),
and relaxing ($z=0.01$) phases.
As we further discuss below, a different initial magnetic structure
leads to a different final strength in the simulated galaxy cluster.
In order to better visualize the spatial differences between our models in
the projected magnetic field distribution, we normalize in
Figure~\ref{fig:projections} the distributions for the Batchelor and
Saffman models by a factor of 10.
These two models, being initially correlated on smaller scales, already reach the
lowest magnetic field strengths at early redshifts, $z\sim10$ (before the cluster forms), and, later on, at all stages of the cluster evolution.

\begin{figure}[t]
    \centering
    \includegraphics[width=8.8cm]{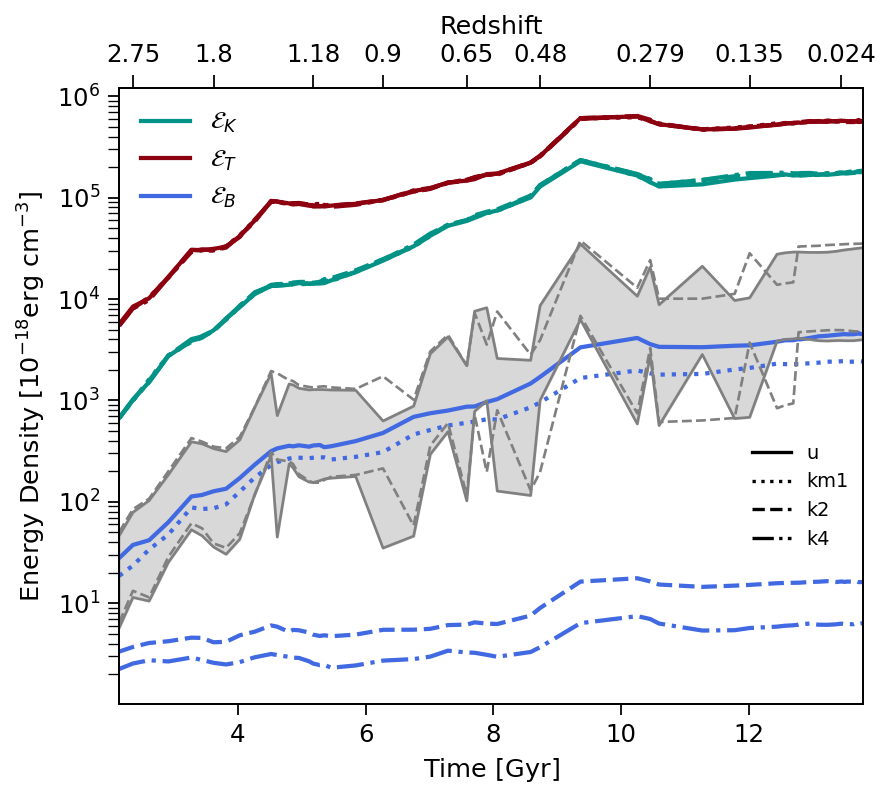}\\
    \caption{Evolutions of the thermal, kinetic, turbulent kinetic, and magnetic
    energy densities, obtained from a comoving box with a side length
    of $1.5\,\cMpch$.
    The solid, dotted, dashed, and dashed-dotted lines correspond to
    the uniform, scale-invariant, Saffman, and Batchelor models, respectively.
    The gray shaded area covers the turbulent energies with smoothing scales between $25$ and $100\,\ckpch$ as indicated by the lower
    and upper gray lines, respectively.
    The solid gray line corresponds to the uniform case, while the dashed
    line corresponds to the Saffman model.
    } 
    \label{fig:energies}
\end{figure}

In Figure~\ref{fig:energies}, we compare the mean magnetic energy
density evolution to the evolution of the thermal, kinetic, and small-scale (turbulent) kinetic
energy densities of the cluster, within a comoving box of side length $1.5\,\cMpch$.
We compute the turbulent energy by filtering out motions at large scales. 
At each component of the 3D velocity, we subtract the mean
velocity, smoothed on two different scales of our selection.
Here, we select $25\,\ckpch$ and $100\,\ckpch$ as the fiducial
smoothing scales \citep[for a more elaborate multifiltering technique
see, e.g.,][]{Vazzaetal2012}.
The magnetic energy density growth in the uniform and scale-invariant
cases is correlated with the growth rates of the thermal and kinetic energy
densities.
For example, the approximate power-law growths of the thermal, kinetic,
and magnetic energies in the redshift range $z=3$--0.65 are found to be
$\sim t^{2.6}$, $t^{3.29}$, and $t^{2.77}$, respectively.
By contrast, the magnetic energy density evolutions of the Batchelor and
Saffman models show less pronounced growth than the aforementioned
trends.
These models evolve as $\sim t^{0.38}$ and $\sim t^{0.1}$, respectively.
In addition, we see that the magnetic energy of the cluster reaches
similar levels to the turbulent energy, at all times, only in the uniform and
scale-invariant models.
Overall, we observe the total growth of the turbulent, kinetic, and thermal
energy densities with respect to $z=3$ as being $\sim 700$, $270$, and $100$,
respectively.
On the other hand, the magnetic energy densities of the uniform,
scale-invariant, Saffman, and Batchelor models grow over the same
$\sim 12\Gyr$ time span by factors of $160$, $130$, $5$, and $3$,
respectively.

\subsection{Radial profiles}
\label{sec:Radgeneral}

%
\begin{figure}[htbp]
    \centering
    \includegraphics[width=9.0cm]{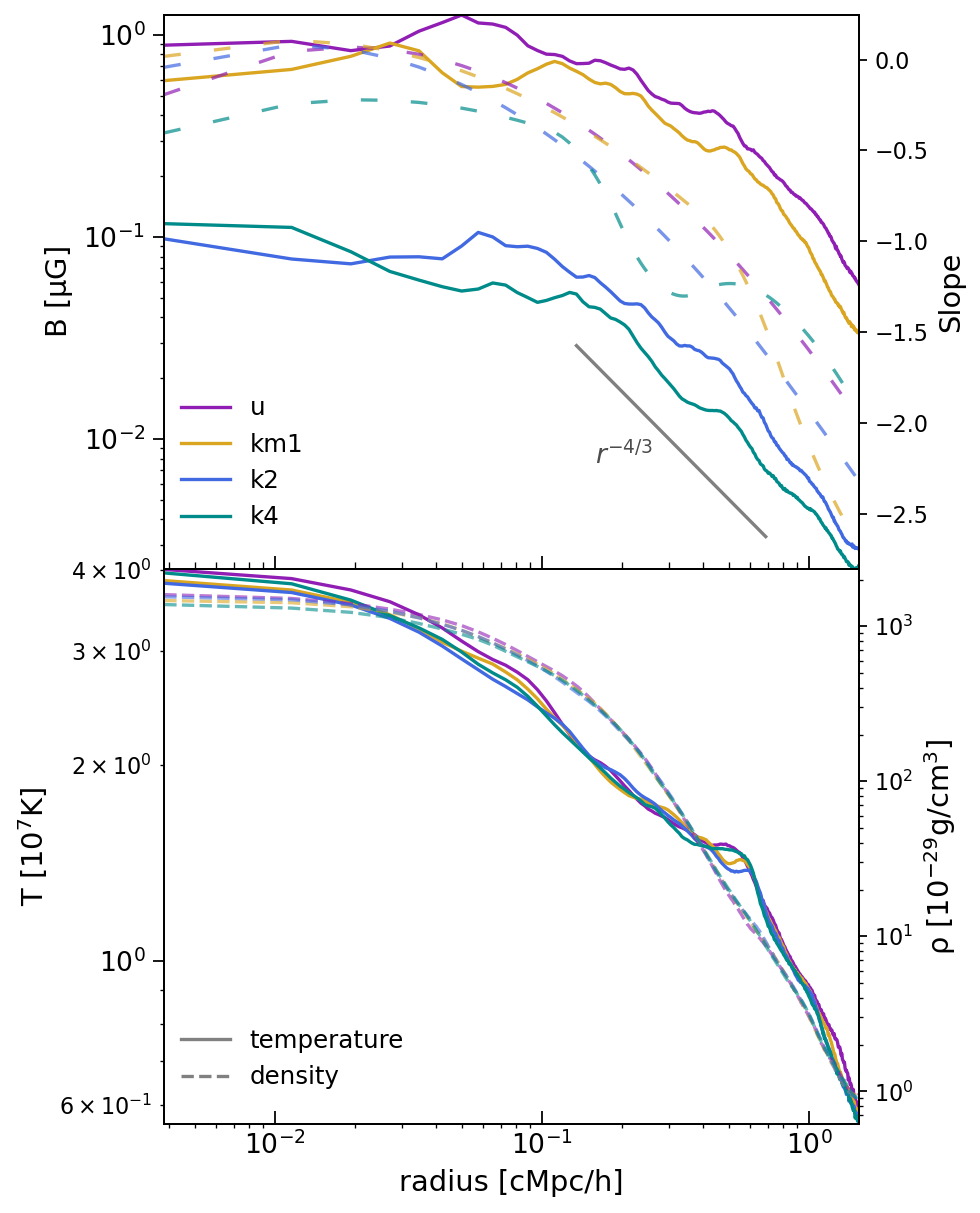}
    \caption{Radial profiles of the magnetic field (top) with the corresponding linear fits (dotted lines) for each magnetic seeding model, and density and temperature fields (bottom).
    All profiles are calculated in a sphere with $r=R_{\text{vir}}$ radius.
    In the outskirts, 
    the magnetic field scales as $r^{-1.19}, r^{-1.39}, r^{-1.5}, r^{-1.34}$ for the uniform, scale-invariant, Saffman, and Batchelor models, respectively.
    } 
    \label{fig:BTRho_radial}
\end{figure}

The radial profiles of our cluster are shown in Figure~\ref{fig:BTRho_radial}.
In the top panel, we show the magnetic field profiles, along with the
expected trend from adiabatic flux freezing ($\propto r^{-4/3}$) and
the slope profiles.
As previously mentioned, we observe that those initial conditions
with more magnetic power at large scales, such as the uniform and
scale-invariant models, show the largest field strengths.
Conversely, as shown in the bottom panel of Figure~\ref{fig:BTRho_radial},
neither in the trends of the slope nor in the radial temperature and
density profiles do we observe any significant differences.

A commonly used proxy for relating the magnetic field and density
distributions is combining their radial dependencies.
In the outskirts ($r>150 \ckpch$), this leads to
$B_{\text{uni}} \propto \rho^{0.43}$, $B_{\text{inv}} \propto \rho^{0.50}$, $B_{\text{Saff}} \propto \rho^{0.54}$, and
$B_{\text{Batch}} \propto \rho^{0.49}$ for the studied models.
These trends are similar to those inferred from the radio observations of the massive $M_{200} \sim 1.8 \times 10^{15} M_{\odot}$
\citep[][]{Kuboetal2007} Coma cluster \citep[][]{Bonafedeetal2010},
but are smoother than the slopes that have been found, e.g., in the observations
of the less massive cluster $M_{200} \sim \times 10^{14} M_{\odot}$ \citep[][]{Girardietal1998} A194 \citep[][]{Govonietal2017}.
It should also be noted that the strength of the magnetic field in the core of the Coma cluster has been found to be higher \citep[$4.7 \uG$;][]{Bonafedeetal2010} than the
obtained values from our simulations.
This can be explained by the fact that the simulated galaxy cluster in
our work is still dynamically young \citep[see, e.g.,][who find that
dynamically older relaxed clusters have larger magnetic field strengths
in the ICM]{Xuetal2011}.
In general, we find these trends to be in good agreement with the results
of \cite{Vazzaetal2018} and \cite{Dominguezetal2019}, where the authors
having studied the dynamo amplification in the simulated galaxy clusters, also using AMR.

\begin{figure*}[htbp]
    \centering
    \includegraphics[width=18.0cm]{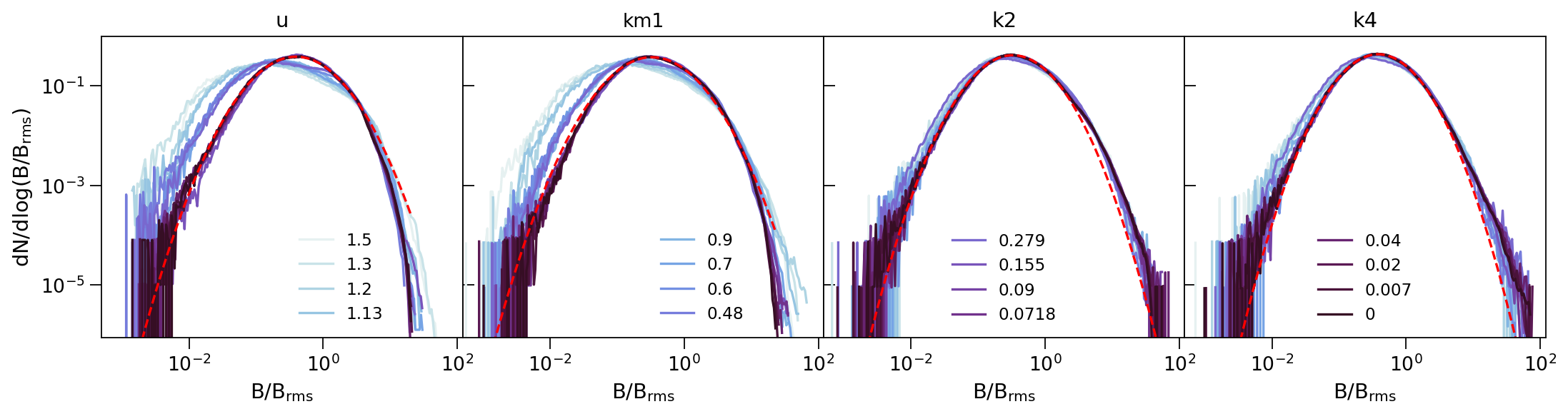}
    \caption{Redshift evolution of the PDFs. From left to right: the uniform, scale-invariant, helical, and nonhelical seedings.
    The PDFs are obtained within a sphere having $R_{\text{vir}}$ radius. The dashed red lines show the lognormal fits for each model.
    } 
    \label{fig:BfieldPDF}
\end{figure*}
%

\subsection{Probability distribution function and curvature}
\label{sec:PDFs}

The distribution of the magnetic fields has been the subject of several works.
It follows from the induction equation that in the diffusion-free
regime and at the kinematic stage of the dynamo (the weak-field limit),
the magnetic field is characterized by a lognormal probability distribution function \citep[PDF; see,
e.g.,][]{Choetal2002, Schekochihinetal2002c, Schekochihinetal2004a,
BranSub2005}.
The lognormality of the magnetic PDFs
is qualitatively understood in terms of the central limit theorem, which is applied to the induction equation (without the diffusion term).
A more rigorous derivation of this result involves the Kazantsev-Kraichnan
dynamo model \citep[][]{Kazantsev1968,Kraichnan1967}.
Following this model, it is possible to predict the evolution of
the mean and the dispersion \citep[see, e.g., Equations~(5) and (6)
in][]{Schekochihinetal2002c} of the lognormal distribution of the
magnetic field.
The spread of the PDF of $\log B$ at bobth the low and high tails of
the distribution is an important characteristic of a lognormal distribution, meaning that a fluctuating magnetic field possesses a high
degree of intermittency, i.e., the fluctuations tend to
become more sparse in time and space and on smaller scales (see,
e.g., \citealt{BeresLaz2019}). In the saturated state of the dynamo, this
intermittency is partially suppressed, and the PDF develops an exponential
tail (see, e.g., \citealt{Schekochihinetal2004a} and the recent simulations
of \citealt{SetaFed2020}).

In the following, we check whether dynamo action is present in our
simulations.
A comprehensive criterion for dynamo action in the presence
of gravity is still missing; see \cite{BN22} for some attempts.\footnote{We
refer here to the earlier papers by \cite{Suretal2010, Suretal2012,
Schoberetal2012, McKeeetal2020, XuLaz2020}, who study the turbulent
dynamo in the context of the formation of the first stars.}
We follow the diagnostics presented in \cite{Schekochihinetal2004a} which have also been used in \cite{Vazzaetal2018} and \cite{Steinwandeletal2021}.

In Figure~\ref{fig:BfieldPDF}, we show the evolution of the 
normalized magnetic field ($B/B_{\text{rms}}$) PDF for all four models.
In the kinematic stage of the dynamo, \cite{Schekochihinetal2004a} find that
the magnetic PDF converges onto a single stationary profile, which is referred to as the self-similarity of the field strength.
In our simulations, we find that the PDFs of the Saffman and Batchelor models
resemble the stationary profile, while the large-scale models (uniform
and scale-invariant) do not show the same behavior toward the low end
tail of the PDF.
The dispersions of the PDFs in the latter two cases decrease (although
not significantly), while the dispersions of phase transition-generated models remain mostly constant.
At the final redshift, we overplot a lognormal fit in
Figure~\ref{fig:BfieldPDF}, and show that the low- and high-end
tails of the distribution are reasonably well fitted by a lognormal
distribution for all PMF models.
Finally, we compute the kurtosis at $z=0$ and obtaine the
values $12$, 13, 31, and $68$ for the uniform,
scale-invariant, Saffman, and Batchelor models, respectively.
These values confirm that all our models exhibit super-Gaussian profiles.

\begin{figure}[htbp]
    \centering
    \includegraphics[width=8.7cm]{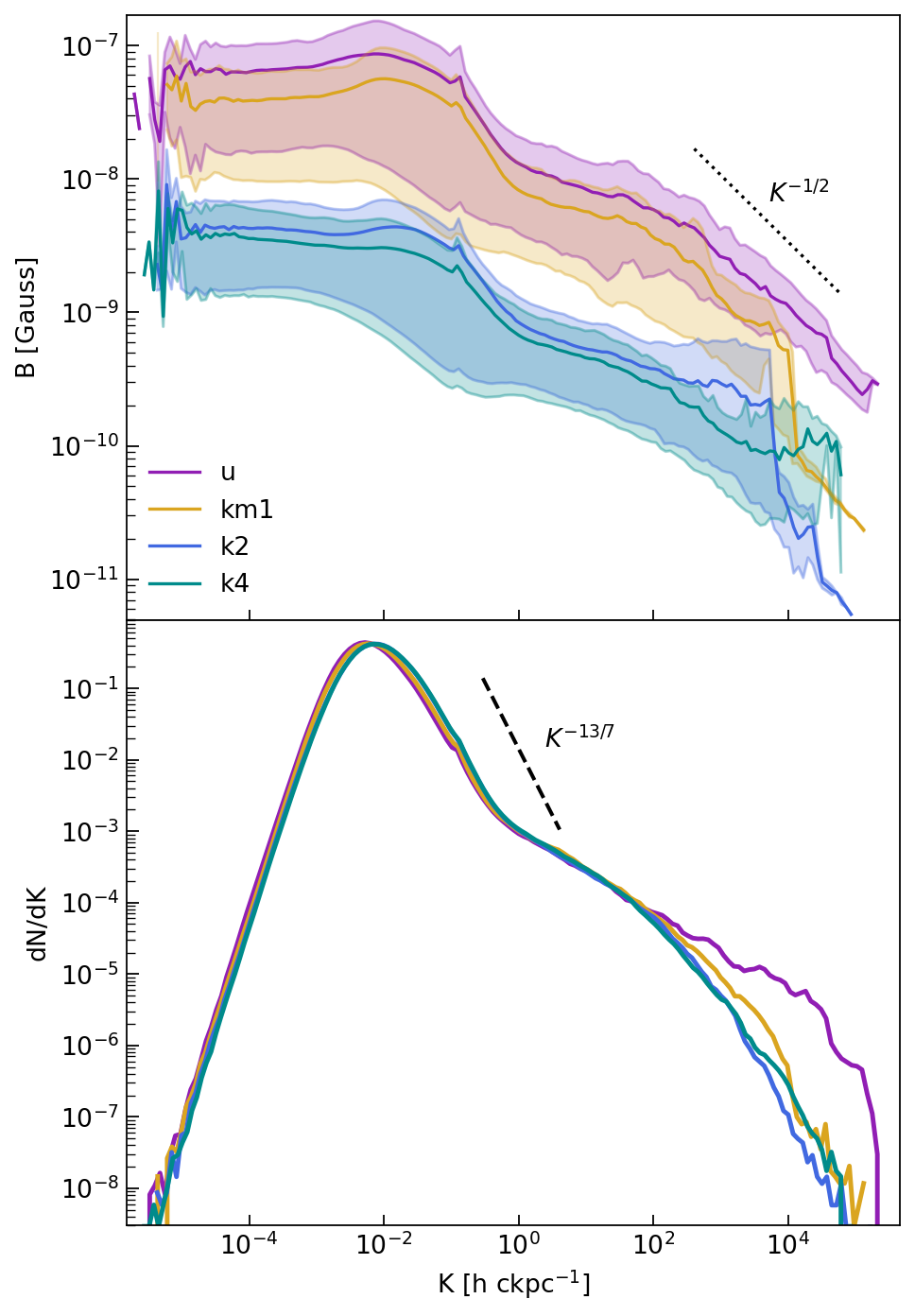}
    \caption{
    Profile of the magnetic field vs. curvature ($\mathbf{|K|}$)
    and the curvature PDFs calculated from a $(3.0 \cMpch)^3$ box at $z=0$.
    The dotted and dashed lines in the panels indicate the scalings that are expected
    from theoretical estimations \citep[from][]{Schekochihinetal2004a}.
    The shaded regions for each model cover the distribution points between
    the 16th and 84th percentiles.}
    \label{fig:Curv}
\end{figure}

The geometry of the magnetic field lines can be studied in terms of the
curvature $\mathbf{K}$ defined as \citep[e.g.][]{Schekochihinetal2002b}:
\begin{equation}
\label{eq:curvature} 
\mathbf{K} = \frac{(\mathbf{B} \cdot \nabla )\mathbf{B}}{|\mathbf{B}^2|} =
\frac{1}{\mathbf{B}^2}\bigg[ \frac{1}{2} \nabla \mathbf{B}^2
- \mathbf{B} \times (\nabla \times \mathbf{B}) \bigg].
\end{equation}
In Figure~\ref{fig:Curv}, we show the dependence of the magnetic field on
the absolute curvature, $K= |\mathbf{K}|$ (top panel) and the curvature
distribution (bottom panel), at $z=0$.
In small-scale dynamo theory, the turbulent amplification of the field
proceeds by the stretching and bending of field lines by turbulent
eddies, resulting in folded structures \citep[see, e.g., Figures 1 and 2 of][]{Schekochihinetal2002c}.
Due to flux conservation arguments, it is expected that the magnetic
field strength will be larger in the stretched segments of field lines,
while the strength will remain small in the bends---i.e., the field
strength and its curvature are expected to be anticorrelated.
This is similar to an earlier finding that stronger flux tubes are
also straighter \citep{BPS95}.
The top panel of Figure~\ref{fig:Curv} presents a good illustration of this
hypothesis.
We observe a declining profile of the magnetic field strength with
increasing curvature of the field.
This anticorrelation is confirmed by calculating the correlation
coefficient between the curvature and the magnetic field $C_{K,B}$
(see Equation~(26) in \citealt{Schekochihinetal2004a}).
For all our models at $z=0$, we obtain $C_{K,B} \sim -0.999$, which is
practically its minimum possible value.
We also note that this anticorrelation has already been observed from earlier redshifts in our simulations. At $z=0$, we obtain the slopes: $-0.32$
($-0.46$), $-0.42$ ($-0.39$), $-0.35$ ($-0.47$), and $-0.25$ ($-0.34$)
for the $(1.5 \cMpch)^3$ region ($(3 \cMpch)^3$ region), corresponding to
the uniform, scale-invariant, Saffman, and Batchelor models, respectively.
Another interesting feature that we see in the top panel of Figure~\ref{fig:Curv} is the flattening of the magnetic field profile toward extremely low
curvatures.
From the bottom panel of Figure~\ref{fig:Curv}, we see that this happens
for $K \lesssim 7 \times 10^{-3} \ckpchI$, where we observe a steep
decrease ($\sim K^{2.5}$) in the curvature PDFs.
The bulk of the curvature distribution is concentrated at the peak values
corresponding to the $192$, 175, 140, and $143\ckpch$ scales\footnote{We
note that our definition of the curvature scale is different from the
definition adopted in \cite{ChoRyu2009}.} (henceforth referred to
as the curvature scales, $\lambda_K$) for the uniform, scale-invariant,
Saffman, and Batchelor models, respectively.
These scales reflect the typical bending scale of the field lines.
As we shall see in Section~\ref{sec:CharScales}, $\lambda_K$ is comparable
to the scale containing the largest magnetic energy.
We find that the peaks of the curvature PDFs shift to the right for all
our models during the major merging phase, i.e., $\lambda_K$ decreases.
This shows that mergers tend to further compress the existing folded
structure, rather than elongating it.
Finally, we also observe a distinctive difference between the uniform
and stochastic models, with the former exhibiting the largest curvatures.

In summary, all of the PMF scenarios
attain intermittent structures (the lognormality of the PDFs) during their 
evolutions even though the growth of the magnetic energy is relatively lower
for the Saffman and Batchelor models (see Figure~\ref{fig:energies}).
(2) There is an anticorrelation between the field strength and the
curvature for all models; however, the curvature scales
are different for the large- and small-scale correlated fields. 
As a result, the different growth rates of the PMFs---i.e., the possible
suppression or excitation of the dynamo---may leave imprints on the
scale, where the further stretching and bending of the field lines is
counteracted by the stronger fields.

\subsection{Spectral evolution} 
\label{subsec:MagPS}

%
\begin{figure*}[htbp]
    \centering
    \includegraphics[width=18.0cm]{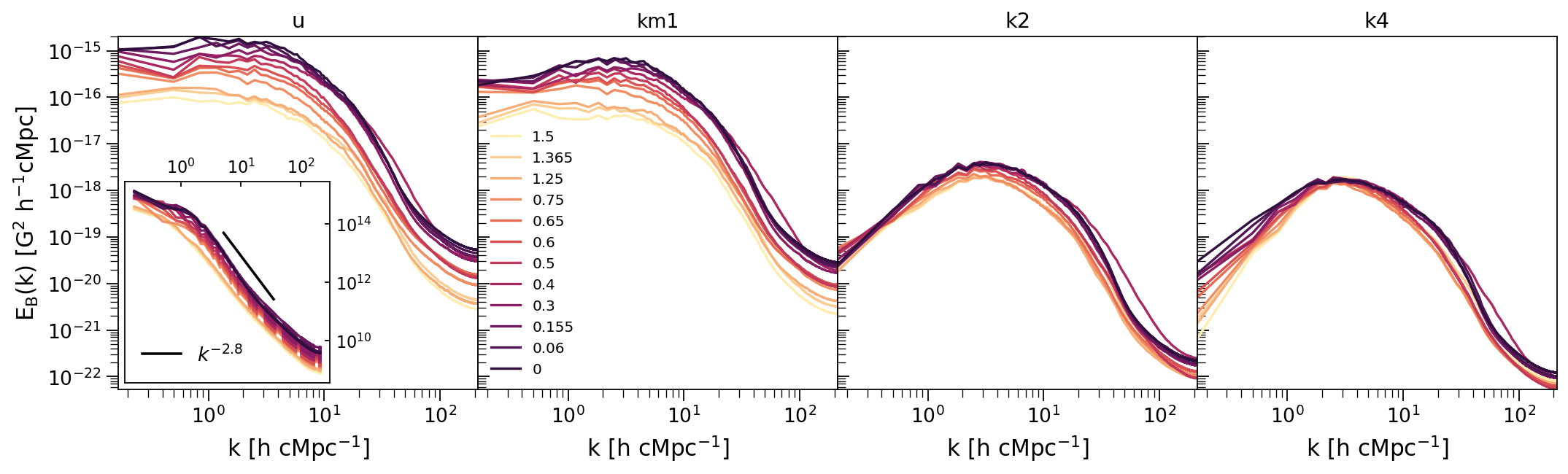}
    \caption{Redshift evolutions of the magnetic and the kinetic
    (inset in the left panel) energy spectra.
    From left to right: the uniform, scale-invariant, Saffman, and
    Batchelor models.
    The energy spectra are calculated from the $(3.0 \cMpch)^3$ box at
    the seventh level of AMR, using the \textit{yt} interpolation method \citep{yt-Turk2011}.
    For additional effects on the shapes and amplitudes
    of the magnetic energy spectra, we refer the reader to
    Appendix~\ref{App:ResTest}.
    The axis units in the inset are $\cm^2 \s^{-2} \cMpch$
    and $\cMpchI$, for the 
    kinetic energy and wavenumbers,
    respectively.
    }
    \label{fig:evBPS}
\end{figure*}

In observations, previous knowledge of the magnetic energy spectrum
is required, in order to obtain more information about the general characteristics
of the magnetic fields in the ICM \citep[see, e.g.,][]{Murgiaetal2004,
Govonietal2006, Govonietal2017, Stuardietal2021}.
The power spectrum of the magnetic field is defined as the Fourier
transform of the magnetic field's two-point correlation function,
$\langle \mathbf{B}_{i}(\mathbf{x})\mathbf{B}_{j}(\mathbf{x}+\mathbf{r}) \rangle$,
where the angle brackets denote the ensemble average and
$\hat{r}=r_i/|\mathbf{r}|$ (see \citealt{MoninYaglom1971} or
\citealt{Brandenburgetal2018}, and references therein).
In practice, we define the magnetic energy power spectrum
$E_B(k)$ through:
\begin{equation}
\label{eq:BPS}
\int E_B(k) dk=\frac{1}{2V}\int \hat{\textbf{B}} \cdot \hat{\textbf{B}}^* 4\pi k^2 dk,
\end{equation}
where $\hat{\textbf{B}}$ denotes the Fourier transform of the magnetic field,
with $\hat{\textbf{B}}^*$ being its complex conjugate, $k=|\textbf{k}|$ is the norm of
the wavevector, and $V$ is the volume that normalizes the spectrum. 

In Figure~\ref{fig:evBPS} we show the evolutions of the magnetic
energy spectra of our four models, with a specific kinetic energy spectrum
for the uniform model being shown in the inset of the first panel.
The magnetic spectrum is computed using Equation~\eqref{eq:BPS} for
different time snapshots, in a $(3 \cMpch)^3$ simulation box, which follows the cluster center as it evolves.
From the figure, one can see that differences between the 
spectra of the 
inflation- and phase transition-generated seed fields arise in both the amplitudes and the shapes of the magnetic power spectra. 
The differences observed in the shapes 
are more pronounced toward the largest scales ($\gtrsim 0.5 \cMpch$) of the simulated galaxy cluster.  In particular, at these scales, the spectra corresponding to the uniform and scale-invariant models 
are flatter than the spectra corresponding 
to the Saffman and Batchelor models. 
A similar result has also been found in Paper~I.
We will further discuss the shape of the magnetic energy spectrum in
Section \ref{sec:fitting}, where we parameterize our four cases. 
On the other hand, we note that the kinetic energy spectra
(the inset in the left panel of Figure~\ref{fig:evBPS}) of our
simulations do not show differences between different PMF models.
The spectra follow a $k^{\delta}$ profile, where $\delta$ changes
between $\sim -2.3$ and $-2.8$ at small scales ($\lesssim 0.5 \cMpch$) over the $9.5 \Gyr$ time span.

In order to understand the differences in the magnetic field amplitudes
between the different models, we recall that at early times
($10 \lesssim z \lesssim 50$), before the cluster forms, only the uniform field model shows amplification homogeneously on all
scales (see Figure~6 of Paper~I),\footnote{A similar result has also been
shown by \cite{SetaFed2020}, where the authors found that even in the case
of a nonactive small-scale dynamo, a uniform seed magnetic field is still
linearly amplified, due to the tangling of the large-scale field (see also
the discussion in the Appendix of \citealt[][]{Setaetal2018} and Paper~I).
We remind the reader that
in this latter work, and generally in small-scale dynamo
studies, contrary to the cosmological simulations, the magnetic and
velocity spectra are concentrated at the same scales.} 
 i.e., in the
absence of gravitational accretion and induced turbulent motions, the
stochastic models mostly stay frozen in or show an insignificant decay.
At late times, as the cluster forms, the large-scale stochastic (i.e., the scale-invariant) model shows a similar trend as the uniform
model and the amplitude of the power spectrum grows on all scales.
This happens because the magnetic power is concentrated on the largest
scales, similar to the power corresponding to the density and velocity
fields (this can be seen in Figure~\ref{fig:B-PS_inits} in which we show our selected initial density and velocity power spectra, as well as in the inset in the first panel of Figure~\ref{fig:evBPS}).
In addition, when turbulence develops, it first produces large-scale
eddies that stretch and bend the field lines of those models where the
large-scale magnetic component is present.
In the stochastic small-scale models, magnetic amplification happens
after turbulence cascades down to scales comparable to the corresponding
magnetic coherence scales.
Therefore, the magnetic energy of these models (Saffman and Batchelor)
is prone to less efficient and slower growth.
Furthermore, as \cite{Schekochihinetal2002b} have pointed out, a chaotically
tangled field will decay toward a folding state at a rate comparable to
the rate of the magnetic energy growth.
Thus, the initial slower growth in the Saffman and Batchelor models
will further suppress the folding of the field lines, leading overall to a lesser amplification degree in these models.

We note that the different growth rates (see also Figure~\ref{fig:energies}) for large- and smaller-scale magnetic fields obtained in our simulations are at odds with the results of driven-turbulence simulations; see e.g., \cite{Choetal2009} and \cite{SetaFed2020} who compare the evolutions of uniform (imposed) and random (stochastic) fields in incompressible and compressible 
MHD turbulence settings, respectively. 
Nonetheless, these authors also find a delay in the onset of the linear growth for low initial field strengths \citep[the uniform field case;][]{Choetal2009} or a decay during the initial transient phase  \citep[the random field case;][]{SetaFed2020}. In the latter work, the uniform model does not decay, and it shows rapid growth during this phase; this trend is similar to the results presented in our work.
Contrary to the results of driven-turbulence MHD simulations \citep[see, e.g.,][]{Schekochihinetal2004a,Brandenburgetal2015}, our study does not clearly indicate forward or inverse cascading either. However, we must bear in mind that the ICM is a complex system, in which mergers might alter the aforementioned trends that we have discussed above.

\subsubsection{Characteristic scales}
\label{sec:CharScales}

A clearly visible characteristic of the magnetic energy spectrum is
the peak scale $L_{E_B(k)}$ corresponding to $1200$ and $400 \ckpch$
for the uniform and scale-invariant models, respectively, and to
$316 \ckpch$ scales for the Saffman and Batchelor models.
To determine the largest energy-containing scale of the magnetic field
\citep[see the definition in][]{ChoRyu2009}, we also calculated the peak
scale of $kE_{B}(k)$, i.e, the peak scale of the spectral energy
per mode.
We find similar values of $L_{kE_B(k)}$ for all our models:
$222 \ckpch$ for the uniform and scale-invariant models and 
$171$ and $154 \ckpch$ for the Saffman and Batchelor models. 
We also find that the peak scales of the density, $L_{kP_\rho(k)}$, and velocity, $L_{kE_v(k)}$ spectral energy per mode are the same: $\sim 857 \ckpch$. In the inflationary and phase-transitional models, $L_{kE_B(k)}$ is $\sim$ one-fourth and $\sim$ one-fifth of $L_{kE_\rho(k)}$ and $L_{kE_v(k)}$, respectively.
A similar result has also been found in the MHD simulations of
\cite{ChoRyu2009} where the authors find a $\sim 1/5$ ratio at the saturation
between $L_{kE_B(k)}$ and the driving (injection) scale of turbulence.\footnote{
See also \cite{Kriel+22} and \cite{BRS22}, who studied the dependence of
different characteristic scales on the magnetic Prandtl number.}
Therefore, our results suggest that most of the magnetic energy resides
on scales that are smaller than the gravity-induced scale or the peak scale of
the density and velocity power spectra.

\begin{figure}[t]
    \centering
    \includegraphics[width=8.5cm]{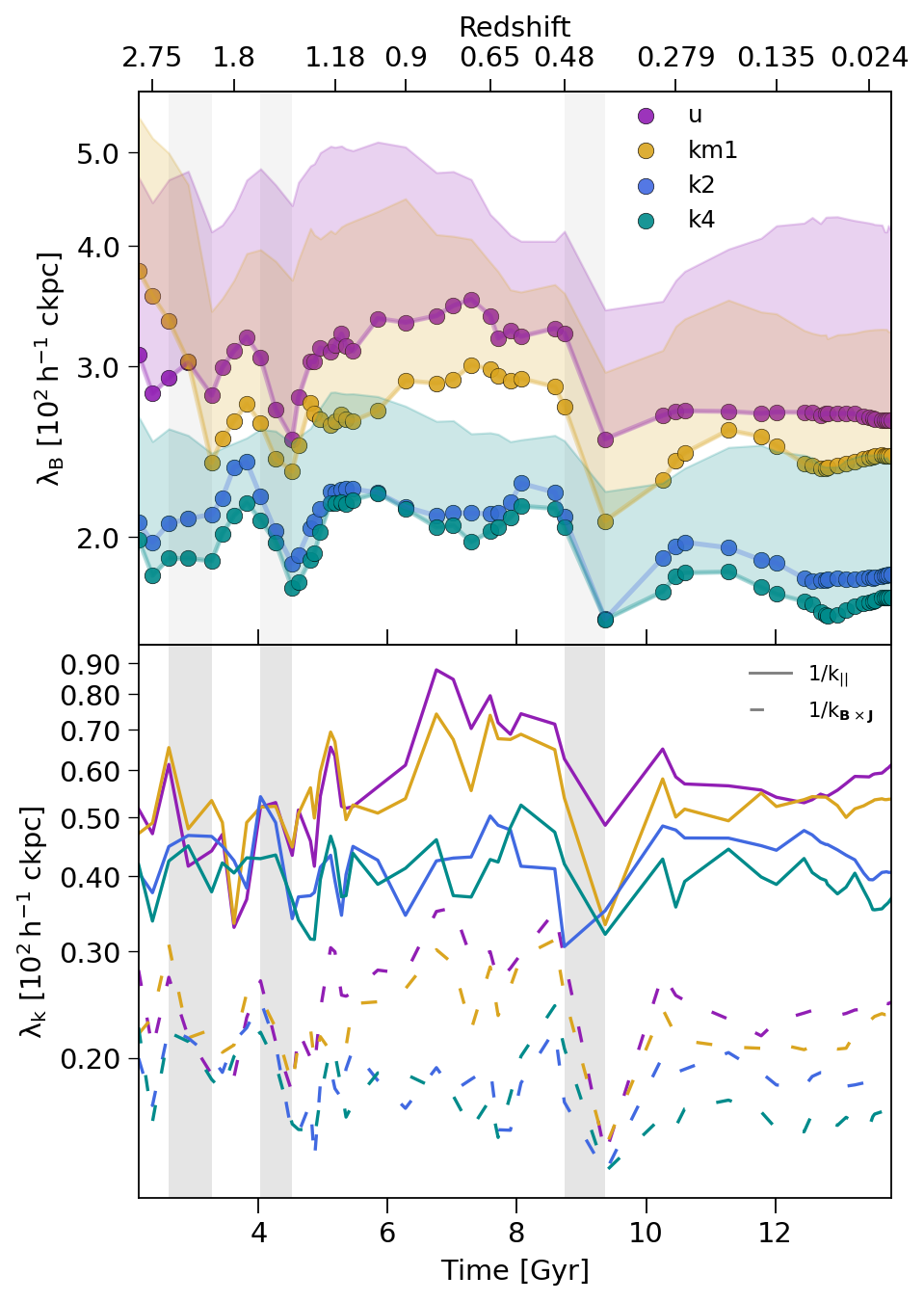}
    \caption{Evolutions of magnetic correlation lengths (top panel) and characteristic parallel and perpendicular scales (bottom panel) for the simulated galaxy cluster. 
    The vertical shaded regions show merging phases during the evolution of the galaxy cluster.
    The horizontal shaded areas in the top panel are delimited
    according to the analyzed region; the lower (upper) lines correspond to
    a $(1.5 \cMpch)^3$ ($(3.0 \cMpch)^3$) region.
    }\label{fig:CorLen}
\end{figure}

The correlation length, which is also referred to as the coherence or integral scale, of the magnetic field is defined as:
\begin{equation}
    \lambda = \frac{\int_0^\infty dk\, k^{-1} E_B(k,t)}{ \int_0^\infty dk \, E_B(k,t)}.
\label{eq:CorLength}
\end{equation}
The evolutions of the magnetic correlation lengths for the different PMF models are shown in the top panel of Figure~\ref{fig:CorLen}. 
We computed the correlation length throughout the $12$ Gyr period,
focusing on a $(1.5\,\cMpch)^3$ region (as in Figure~\ref{fig:energies}).
We also conducted the same analysis in a $(3.0\,\cMpch)^3$ region, since the correlation length can depend on the box size under consideration.
During merger events (shown as the vertical shaded areas in Figure~\ref{fig:CorLen}), 
the magnetic correlation length decreases for all four models.
This happens mainly because compression becomes dominant
as the infalling gas clump crosses the cluster center.\footnote{We note that merger events add additional power as they enter the analyzing box; therefore, this can also contribute to the decrease of the magnetic correlation length.}
The same effect has also been observed in other cosmological MHD simulations, e.g. in \cite{Dominguezetal2019}, where the authors find that major merger events shift the magnetic power toward smaller scales. 
It is after each merger event that the magnetic correlation length increases again
for all four models.

Finally, as the cluster enters its relaxing phase at
$z\lesssim 0.135$, the correlation lengths for all models converge to
$260$--$410$, $240$--$330$, $180$--$230$, and $170$--$240\ckpch$ for the uniform,
scale-invariant, Saffman, and Batchelor models, respectively.
These values are one order of magnitude larger than those that are obtained and
typically referred to as the coherence scale (a few tens of kiloparsecs) from radio
observations \citep[see, e.g.,][]{Murgiaetal2004,VogtEnsslin2005,Weerenetal2019}.
The strongest differences in the magnetic correlation lengths between the models are 
better seen at earlier redshifts, where the scale-invariant model shows a coherence length that is 
larger than those of the Saffman and Batchelor models by a factor of $\sim 2$. 
We note that while the differences between the uniform and scale-invariant models and those between the Batchelor and Saffman models decrease after the merger events, we still observe larger correlation lengths in the inflationary cases than in the phase-transitional scenarios throughout the evolution of the galaxy cluster over this 12 Gyr period.

Following \citealt{Schekochihinetal2004a}, 
one can also define the characteristic wavenumbers,
\begin{equation}
\begin{gathered}
k_{\parallel} = \bigg( \frac{\langle |\mathbf{B \cdot \nabla \mathbf{B}}|^2\rangle }{ \langle B^4 \rangle} \bigg)^{1/2},~~ 
k_{\mathbf{B}\times \mathbf{J}} = \bigg( \frac{\langle |\mathbf{B \times \mathbf{J}}|^2\rangle }{ \langle B^4 \rangle} \bigg)^{1/2}
\end{gathered}
\end{equation}
corresponding to the magnetic field variation along ($k_{\parallel}$)
and across ($k_{\mathbf{B}\times \mathbf{J}}$) itself, with $\mathbf{J}$
being the current density.
In small-scale dynamo theory, it has been argued that generally
$k_{\mathbf{B}\times \mathbf{J}} > k_{\parallel}$ since the shear
flows can more rapidly stretch and reverse the field lines in the
plane transverse of the field line itself \citep[see][and references
therein]{Schekochihinetal2002b}.
In other words, the growth of the typical fluctuation wavenumber
$k=\sqrt{k_{\mathbf{B}\times \mathbf{J}}^2 + k_{\parallel}^2}$ should mostly be due to the increase of $k_{\mathbf{B}\times \mathbf{J}}$.
It has been shown that in both the MHD dynamo
\citep[][]{Schekochihinetal2004a} and the plasma dynamo
\citep[][]{St-ongeetal2018}, 
the $k_{\mathbf{B}\times \mathbf{J}} > k_{\parallel}$
ordering is satisfied
in the initial, rapid growth phase and that is persists in the 
kinematic and
nonlinear regime of a 
dynamo (during saturation).

In the bottom panel of Figure~\ref{fig:CorLen} we show the evolution
of the $\lambda_{\parallel}, \lambda_{\mathbf{B}\times \mathbf{J}}$,
scales corresponding to the inverse
$k_{\parallel}$, $k_{\mathbf{B}\times \mathbf{J}}$,
characteristic wavenumbers, respectively.
The condition $k_{\mathbf{B}\times \mathbf{J}} > k_{\parallel}$
is satisfied for $z<3$ in the simulated cluster for all four magnetic cases.
We find a maximum ratio of $k_{\mathbf{B}\times \mathbf{J}}/ k_{\parallel} \sim 2-3$ over the 12 Gyr period.
The ordering of these characteristic scales seems to be
consistent with the arrangement of a magnetic field in
folded structures;
see also Figure~23(a) of \cite{Schekochihinetal2004a}.
This result, along with the lognormarlity of the PDFs and curvature results, would be compatible with the kinematic stage of a dynamo in
our simulations.

\subsubsection{Parameterization of magnetic energy spectra}
\label{sec:fitting}

\begin{figure}[t]
    \centering
    \includegraphics[width=8.5cm]{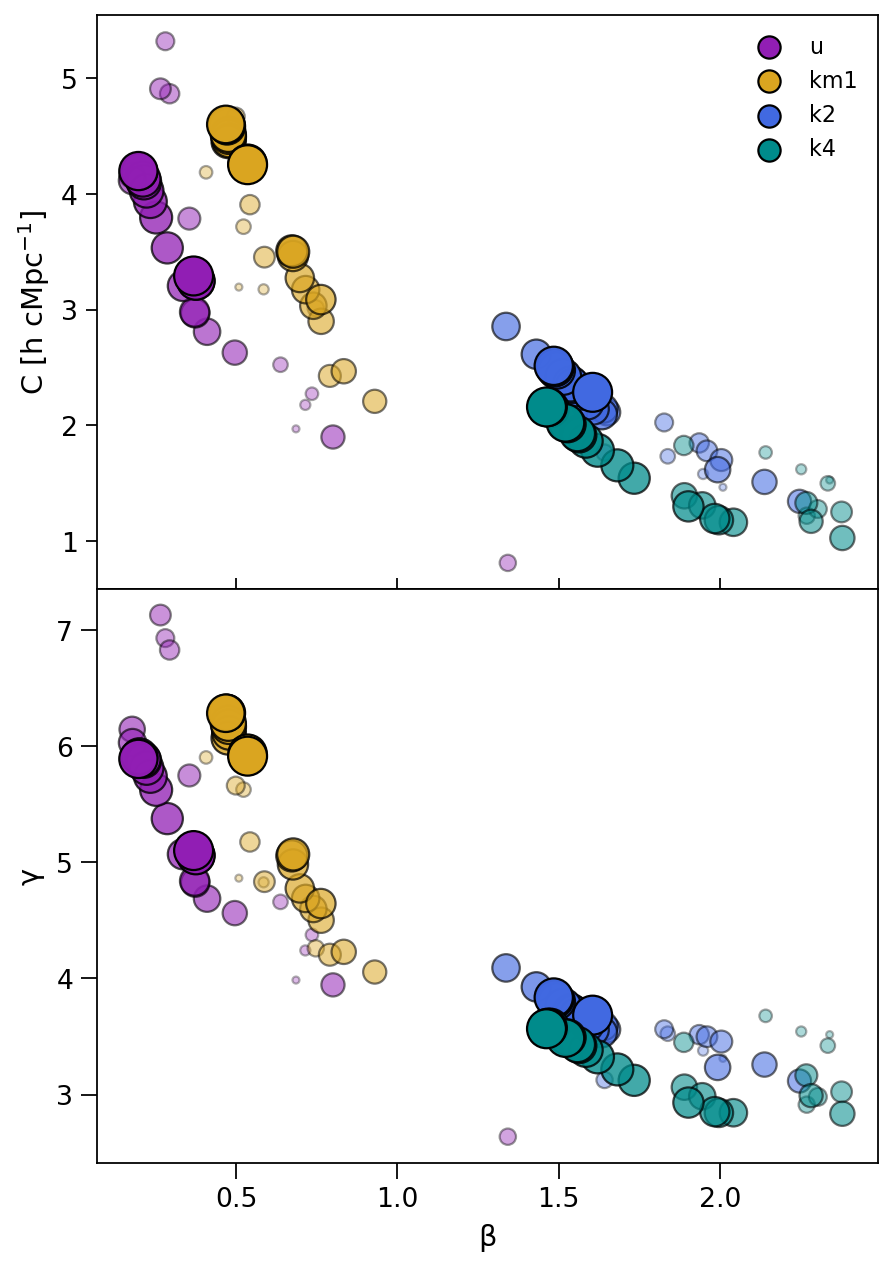}
    \caption{Parameter spaces for the best-fit parameters of our different PMF models considering a $(3.0 \cMpch)^3$ region. The smaller markers and the lower-opacity colors show the parameters at early times. The top and bottom panels show the results from the fits according to Equations~\eqref{eq:fittingPa} and \eqref{eq:fittingAx}, respectively. 
    } 
    \label{fig:fitted_params}
\end{figure}

In order to discriminate among the magnetic field models 
we characterize the magnetic energy spectra 
in the $(3\,\cMpch)^3$ box. We consider two different fitting functions. First, we use the equation
\begin{equation}
E_B(k) = Ak^{\beta} \left\{ 1-\text{erf} \left[ B \, \ln\left(\frac{k}{C} \right) \right] \right\},
\label{eq:fittingPa}
\end{equation}
where $A$ gives the normalization, $B$ is related to the width of the
spectra, $C$ is a characteristic wavenumber of the magnetic field, and
$\beta$ is the slope of the spectrum at small wavenumbers.
This fitting function has been used in \cite{Dominguezetal2019} to study
the evolutions of the magnetic energy spectra for a set of highly resolved
galaxy clusters, assuming  a uniform magnetic field seeding.
The large-scale slope used by the authors satisfies the Kazantsev
\citep[][]{Kazantsev1968, Kulsrud+Anderson92} scaling, $\beta=3/2$.
We use a similar approach, by fitting Equation~(\ref{eq:fittingPa}) to
the magnetic energy spectra of our simulated cluster and obtaining the
best-fit parameters $A$, $B$, and $C$.
In our case, we fix the initial $\beta$ at each time step separately. That
is, as a first step, we determine the large-scale slope of the spectra,
$\beta$, and, as a second step, we fix this value in the fitting equation.

The second fitting function is motivated by the MHD simulations in \cite{Brandenburgetal2017}, where a phase transition-generated magnetic
field has a pronounced peak on the scale of the field generation.
We adopt the following spectral shape \citep{Brandenburgetal2017,
RoperPoletal2022}:
\begin{equation}
E_B(k) = (1+D)^{1/\alpha} E_{\rm m} \frac{(k/k_*)^{\beta}}{[1+D(k/k_*)^{\alpha(\beta+\gamma)}]^{1/\alpha}},
\label{eq:fittingAx}
\end{equation}
where $D$ controls the peak scale, $E_{\rm m}$ is the normalization, $k_*$
is the peak wavenumber, and $\beta$ and $\gamma$ are the slopes at large
($k<k_*$) and small ($k>k_*$) scales, respectively.
The value of $\alpha$ is chosen to be $0.25$, 
to ensure a smooth transition between the spectra on large and small scales.
In this case, $D, E_{\rm m},$ and $\gamma$ are the best-fit parameters obtained.
Figure~\ref{fig:fitted_params} summarizes the results of our fitting
procedure, using Equations~(\ref{eq:fittingPa}) and (\ref{eq:fittingAx}).
We only show only the most important best-fit parameters for each model in
Figure~\ref{fig:fitted_params}, while we provide all the parameters at
$z=0$ in Table~\ref{tab:Tab3}.
In the upper panel, we show the $C-\beta$ parameter space (see
Equation~(\ref{eq:fittingPa})), and in the lower panel we show the
$\gamma-\beta$ parameter space (see Equation~(\ref{eq:fittingAx})).
We show the evolutions of the fitting parameters over a time span of
$6.1\Gyr$ in the redshift range of $0.63 \leq z \leq 0$.
As it can be seen from Figure~\ref{fig:CorLen}, this period encompasses
a major merger event at $z\sim0.48$ and the relaxing phase of the cluster.

\begin{deluxetable}{c c c c c  c }
\label{tab:Tab3}
\tablecaption{
Parameters of the power spectra for different models and for
different fitting functions at $z=0$.
The power spectra are fitted with Equations~(\ref{eq:fittingPa}) and
(\ref{eq:fittingAx}).
The fixed $\beta$ parameters are: $0.37$, 0.54, 1.61, $1.46$ for the
uniform, scale-invariant, Saffman, and Batchelor models, respectively, with $\alpha=0.25$.
}
\tablehead{
\colhead{Model} & \colhead{Eq.} &  \colhead{A} $[\mathrm{G}^2 \cMpch]$  & \colhead{B} &  \colhead{C $[\cMpchI]$ } \\
\colhead{} &\colhead{}&\colhead{$E_{\rm m}$} $[\mathrm{G}^2 \cMpch]$  &\colhead{D} &\colhead{$\gamma$}  
}
\startdata
\multirow{2}{*}{u} & \multirow{1}{*}{\eqref{eq:fittingPa}}  &  $8.92 \times 10^{-16}$  & 2.16  &3.29\\
            & \multirow{1}{*}{ \eqref{eq:fittingAx}}  & $1.63 \times 10^{-15}$ & 0.03  & 5.10 \\
\hline
\multirow{2}{*}{km1} & \multirow{1}{*}{ \eqref{eq:fittingPa}} &  $2.54 \times 10^{-16}$   & 2.56 & 4.25   \\
            & \multirow{1}{*}{ \eqref{eq:fittingAx}}  & $6.44 \times 10^{-16}$  & 0.095 & 5.92 \\
\hline
\multirow{2}{*}{k2} & \multirow{1}{*}{ \eqref{eq:fittingPa}} &  $8.66 \times 10^{-19}$  & 2.27 & 2.29   \\
            & \multirow{1}{*}{ \eqref{eq:fittingAx}}  & $3.57 \times 10^{-18}$  & 0.403 &  3.68 \\
\hline
\multirow{2}{*}{k4} & \multirow{1}{*}{ \eqref{eq:fittingPa}} &  $4.91 \times 10^{-19}$  &  2.17 & 2.16   \\
            & \multirow{1}{*}{ \eqref{eq:fittingAx}}  & $1.62 \times 10^{-18}$  & 0.427 &  3.57 
\enddata
\end{deluxetable}

The $C$--$\beta$ and $\gamma$--$\beta$ parameter spaces highlight how the
spectral characteristics of the inflationary cases differ from those of
the phase-transitional cases.
In the following, we discuss the main points.
\begin{itemize}
    \item[(1)] The evolution of the $C$ parameter 
varies between $2$--$4.5 \, \cMpchI$ for the 
inflationary models and between $1$--$2.8\cMpchI$ for the phase transitional models. 
The ratio between the magnetic correlation length and $1/C$ is $\sim 1.4$
for the inflationary models and $\sim 0.5$ for the phase-transitional
seedings.
That is, $\lambda_B \gtrsim 1/C$ for the former scenarios
and $\lambda_B \lesssim 1/C$ for the latter models.
This shows that this fitting equation is a good proxy for obtaining a
characteristic scale of the magnetic field that can be comparable to or of
the same order as $\lambda_B$.

\item[(2)]The large-scale slopes of the magnetic power spectra
characterized by $\beta$ deviate from a Kazantsev slope in
the inflationary models where $\beta \lesssim 1$.
In contrast, the phase-transitional models are approximately characterized
by a Kazantsev slope at late redshifts.
These models show a scatter in the range $1.2 \lesssim \beta \lesssim 2.5$,
where the slope tends to flatten progressively toward $\sim 3/2$ as
the cluster virializes.

\item[(3)] The small-scale slopes of the magnetic power spectra
characterized by $\gamma$ vary between $3.9$ and $6.5$ in the inflationary
models and $2$ and $4.1$ in the phase-transitional models.
As seen from Figure~\ref{fig:evBPS}, the magnetic energy growth at
scales larger than the characteristic scale is more pronounced in the two
inflationary cases, therefore explaining the larger values of $\gamma$
compared to those from the phase-transitional models.

\end{itemize}

Finally, we note that we refrain from claiming that
the phase-transitional models can corroborate the 3/2 large-scale
slope predicted by the Kazantsev model since, as can be seen from
Figures~\ref{fig:evBPS} and \ref{fig:fitted_params}, this slope can vary throughout the
complex evolution of galaxy clusters. Indeed, the multiple merger events that lead to the final formation
of a cluster already break down one of the most basic assumptions of
Kazantsev theory---i.e., a delta-correlated (in time) velocity field.

\section{Numerical aspects}
\label{sec:NumAsp}

The numerical resolution is an important caveat to the analysis conducted in this work.
Similar simulations presented by \cite{Vazzaetal2014,Vazzaetal2018} have shown
that magnetic fields tend to be more strongly affected by resolution
effects than the velocity field, for example.
Therefore, the growth rates of the seed magnetic
fields in galaxy clusters are also resolution-dependent.
Within our numerical setup, we assess the convergence of our results
by performing extra simulations with different AMR levels.
In Appendices~\ref{App:ResTest} and \ref{App:AMRlevelsProf}, we show
how the power spectra, the PDFs, and the radial profiles of the magnetic
field have already converged at six AMR levels (on scales $\gtrsim50\ckpch$).

As in Paper~I, we rely on the Dedner cleaning algorithm
\citep{Dedneretal2002} to impose the $\nabla\cdot\BB=0$ condition.
While the Dedner formalism has been found to be robust and accurate,
as well as to converge quickly on the right solution for most idealized
test problems \citep[][]{WangAbel2009,WangAbel2010,Bryanetal2014}, and for other more realistic astrophysical applications
\citep[][]{HopkinsRaives2016,Triccoetal2016,Barnesetal2018}, this method may be limited compared to the constrained transport
(CT) schemes \citep[][]{Kritsuk2011}.
The intrinsic dissipation of the Dedner scheme, via cleaning waves, can affect the final magnetic growth of our PMF models.
Divergence cleaning has also been associated with spurious magnetic
helicity production \citep{BranScan2020}.
Consequently, we cannot entirely rule out the possibility numerics
(see also Appendix~C of Paper~I) can  also contribute to the obtained differences between the growth rates of the inflationary and phase-transitional models. 
In Figure~\ref{fig:divergence_radial} of Appendix~\ref{App:ResTest},
we show the radial profile of the magnetic field divergences in our
simulated cluster.
The densest central region of the cluster exhibits a similar normalized divergences for our four PMF models, while some differences between
the inflationary and phase-transitional cases can only be observed at large
radii, $\gtrsim 1.2 \cMpch$, with the former case showing the lowest
values.
Nevertheless, the Dedner cleaning method keeps the numerical magnetic
field divergence below $\sim 5 $\% ($\sim 8 $\%) of the local magnetic
field within the cluster volume having $r=R_{500}$ ($r=R_{100}$) radius.
This shows that the divergence remains reasonably low 
in the largest fraction of the simulated cluster volume.
We leave a numerical comparison between the Dedner and CT schemes within the \texttt{Enzo} code in the context of PMFs 
in galaxy clusters for future work.

As mentioned in Section~\ref{sec:simulations}, we have only focused on the
amplification of PMFs, due to the structure formation and turbulent motions
in the ICM.
However, the inclusion of additional physics, such as feedback and
radiative cooling physics, could lead to larger amplification levels of
our PMF models, and may therefore affect the final magnetic fields \citep[see e.g.,][]{Marinnacietal2015,Vazzaetal2017}.
The effects of these processes on distinguishing between
different magnetogenesis scenarios will also be studied in our future
work.

\section{Conclusions}
\label{sec:Summ}

In this work, we have investigated the evolution of PMFs during the formation of a massive galaxy cluster.
We have studied seed magnetic fields resembling inflation-
and phase transition-generated
nonhelical fields.
In the former case, we have assumed either (1) a uniform, constant
magnetic field or (2) a stochastic field.
The stochastic model is motivated by the pre-recombination evolution of an inflationary seed field (initially having a scale-invariant
spectrum), while the uniform case corresponds to the Mukohyama model.
In the case of phase transition-generated seed magnetic fields, we have studied
stochastic models with initial (3) Batchelor and (4) Saffman spectra.
These magnetic spectra are motivated by the causal generation
and evolution of phase-transitional fields until recombination.

The main results of our work can be summarized as follows.

\begin{itemize} 
\item[1.] \textit{Final amplification}. The amplification of a primordial seed magnetic field in the ICM strongly depends on the initial structure of the magnetic field.
In our simulated galaxy cluster, the inflation-generated uniform and scale-invariant models show more efficient amplification compared to the phase transition-generated Saffman and Batchelor models.
We see that in the former cases the magnetic energy density is
of the same order of magnitude as the turbulent energy budget of the cluster.
In such cases, the magnetic power is concentrated on the largest scales,
similar to the power corresponding to the density and velocity fields.
This leads to more efficient turbulent amplification of these large-scale
models compared to the small-scale phase-transitional seed magnetic fields.

\item[2.] \textit{Radial profiles}.
The radial magnetic field profiles at the final redshift ($z=0$)
reflect the aforementioned differences in the magnetic energy growth.
The amplitude of the uniform and scale-invariant models is one
order of magnitude larger ($\sim 0.8$--$1 \uG$; cluster center) than the amplitude attained by the
phase transition-generated magnetic fields ($\sim 0.1 \uG$).
The declining magnetic field profile toward the outskirts reveals the
largest differences between the uniform ($r^{-1.19}$) and the Saffman
($r^{-1.5}$) models.

\item[3.] \textit{Small-scale dynamo}.
All of our models exhibit a degree of small-scale dynamo amplification, as
hinted at by the lognormality of the magnetic field PDFs and the folded
structures of field lines (i.e., the anticorrelation between the field strength
and curvature 
and the ordering of the characteristic wavenumbers).
Consistent with the previous works \citep[][]{Vazzaetal2018,
Dominguezetal2019, Steinwandeletal2021}, we find that cosmological
MHD simulations do not exhibit a small-scale dynamo that can be
compared one-to-one to the Kazantsev theory.

\item[4.] \textit{Coherence lengths}.
We find that, throughout the evolution, the magnetic correlation length
of the cluster depends on both the initial structure of the seed field
and the merger history.
We find that the inflationary models (initially large-scale correlated PMFs)
will inherently attain larger coherence
lengths than the phase-transitional models, throughout the evolutions of
galaxy clusters.
This trend is even persistent during merger events, where the correlation
length decreases for all models.
At the final redshifts, we observe a factor of $\sim 1.5$ difference
in the coherence scales of the uniform and scale-invariant models versus
the Batchelor and Saffman models.
The correlation lengths calculated from a $[(1.5$--$3)\cMpch]^3$ analyzing
box span in the range: $260$--$410$, $240$--$330$, $180$--$230$, and 
$170$--$240\ckpch$ for the uniform, scale-invariant, Saffman, and
Batchelor models, respectively.

\item[6.] \textit{Spectral characteristics}.
We provide two possible fits for the magnetic energy spectra.
The parameterization of the magnetic energy spectra shows how
phase-transitional and inflationary models can be differentiated.
The large-scale slopes (the $\beta$ parameter; see Section~\ref{sec:fitting})
are smaller ($\lesssim 1$) for the inflationary PMFs, but larger
($1.2 \lesssim \beta \lesssim 2.5$) for the phase-transitional
PMFs, over a time span of $6.1$ Gyr ($0.63 \leq z \leq 0$).
The Batchelor and Saffman models have Kazantsev scaling ($\beta=3/2$)
at the final redshifts, even though these fields are amplified to a
lesser degree.
On the contrary, the small-scale slopes (the $\gamma$ parameter;
see Section~\ref{sec:fitting}) are larger
for the inflationary models ($\gamma \sim 3.9$--$6.5$) than for the
phase-transitional seedings ($\gamma \sim 3.9$--$6.5$).
The $1/C$ scales at the final redshift are $300 \ckpch$, $240 \ckpch$,
$440 \ckpch$, and $460 \ckpch$ for the uniform, scale-invariant, Saffman, and Batchelor models, respectively.

\end{itemize}

In summary, we conclude that the two competing scenarios of primordial
magnetogenesis, inflationary and phase-transitional, can indeed be
distinguished on galaxy cluster scales.
The initial structure of the seed magnetic field affects the
efficiency of the dynamo.
Thus, PMFs do not only leave unique imprints on scales larger than
those of galaxy clusters (Paper~I), but it can also influence
small-scale dynamo action in the ICM.
These signatures are reflected in the magnetic energy power spectrum
and the coherence scale of different models.
An analytical power spectrum of the magnetic field is required
for synthetic RM studies \citep[see the method description in, e.g.,]
[]{Stuardietal2021}, giving us the possibility to constrain the structure
of observed galaxy cluster magnetic fields.
We provide two analytical models that
can readily be used
in observational works \citep[see, e.g.,][for such
examples]{Murgiaetal2004,Bonafedeetal2013,Govonietal2017}.

Finally, since the inflationary models show larger field strengths (both in the centers as well as on the outskirts of the simulated clusters) and coherence scales, 
these may make them better candidates for producing 
e.g., the central cluster 
radio diffuse emission in the form of the ``megahalos'' that have been recently detected with LOFAR \citep[][]{Cucitietal2022}. Megahalos 
fill a volume 30 times larger than do common radio halos. 
This makes them 
interesting objects for unveiling the nature of relativistic electrons and magnetic fields on the outskirts of galaxy clusters.
On the other hand, inflationary magnetogenesis scenarios would 
be also favored
for obtaining the fast magnetic field amplification that is needed to explain the observed diffuse radio emission in high-redshift galaxy clusters \citep{DiGennaroetal2021}.
Deeper observations of megahalos, together with the detailed RM images that will be obtained by future observations with the
Square Kilometre Array (SKA) and the upgraded LOFAR 2.0, will
have the potential to unravel the origins of large-scale magnetic fields.

\section*{Acknowledgement}

We thank Franco Vazza for sharing the initial \texttt{Enzo} setup. 
We appreciate useful discussions with and comments from Sergio Martin-Alvarez,
Rajsekhar Mohapatra, Amit Seta, G\"unter Sigl, and Chiara Stuardi.
We also thank the anonymous referee for the useful comments that have improved
the presentation of the manuscript.
S.M.\ acknowledges financial support from the Shota Rustaveli National
Science Foundation of Georgia (SRNSFG, No.\ N04/46-1), 
the Volkswagen foundation, and the German
Academic Exchange Service (DAAD). 
P.D.F.\ acknowledges financial support from the European
Union's Horizon 2020 program, under the ERC Starting Grant
``MAGCOW'', No.\ 714196 with F.\ Vazza as principal investigator.
A.B.\ acknowledges support from the Swedish Research Council
(Vetenskapsr{\aa}det, grant No. 2019-04234).
Nordita is sponsored by Nordforsk. A.B.\ and T.K.\ acknowledge NASA Astrophysics Theory Program (ATP) grant (No.\ 80NSSC22K0825). T.K. acknowledges support from the SRNSFG (grant No.\ FR/19‐8306).

The computations described in this work were performed using the
publicly available \texttt{Enzo} code (http://enzo-project.org), which is
the product of the collaborative efforts of many independent scientists from
numerous institutions around the world.  Their commitment to open science has helped make this work possible. 
We also acknowledge the \textit{yt} toolkit \citep{yt-Turk2011}, which was used as the analysis tool for our project.
The simulations
presented in this work made use of computational resources on Norddeutscher Verbund f\"ur Hoch- und H\"ochstleistungsrechnen (HLRN),
under project no. hhp00046 and were partially conducted on the
JUWELS cluster at the Juelich Supercomputing Centre (JSC) under
project no. 24944 CRAMMAG-OUT, with P.D.F. as
the principal investigator.
We also acknowledge the allocation of computing resources that was provided by the
Swedish National Allocations Committee at the Center for Parallel
Computers at the Royal Institute of Technology in Stockholm and
Link\"oping. 

\section*{Data Availability}
The derived data supporting the findings of this study are available from the corresponding author upon request.

{\large\em Software:} The source codes used for
the simulations of this study---\texttt{Enzo} \citep{2019JOSS....4.1636B} and
the {\sc Pencil Code} \citep{JOSS}---are freely available
online, at  \url{https://github.com/enzo-project/enzo-dev}
and \url{https://github.com/pencil-code/}. \\

\typeout{} 
\bibliography{PMFs}{}
\bibliographystyle{aasjournal}

\appendix

\section{Resolution tests and Divergence }
\label{App:ResTest}

%
\begin{figure}[htbp]
    \centering
    \includegraphics[width=8.cm]{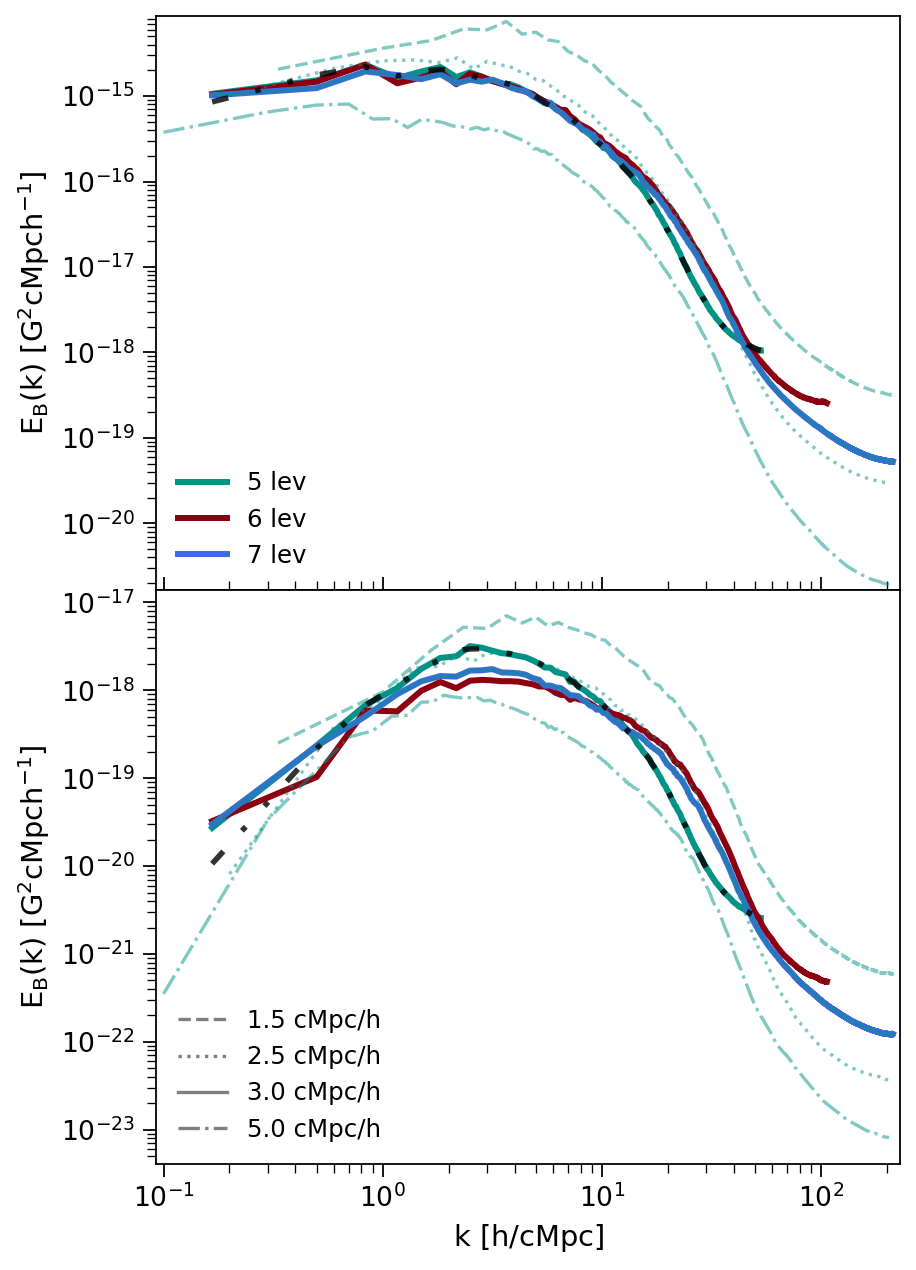}\\
    \includegraphics[width=8.cm]{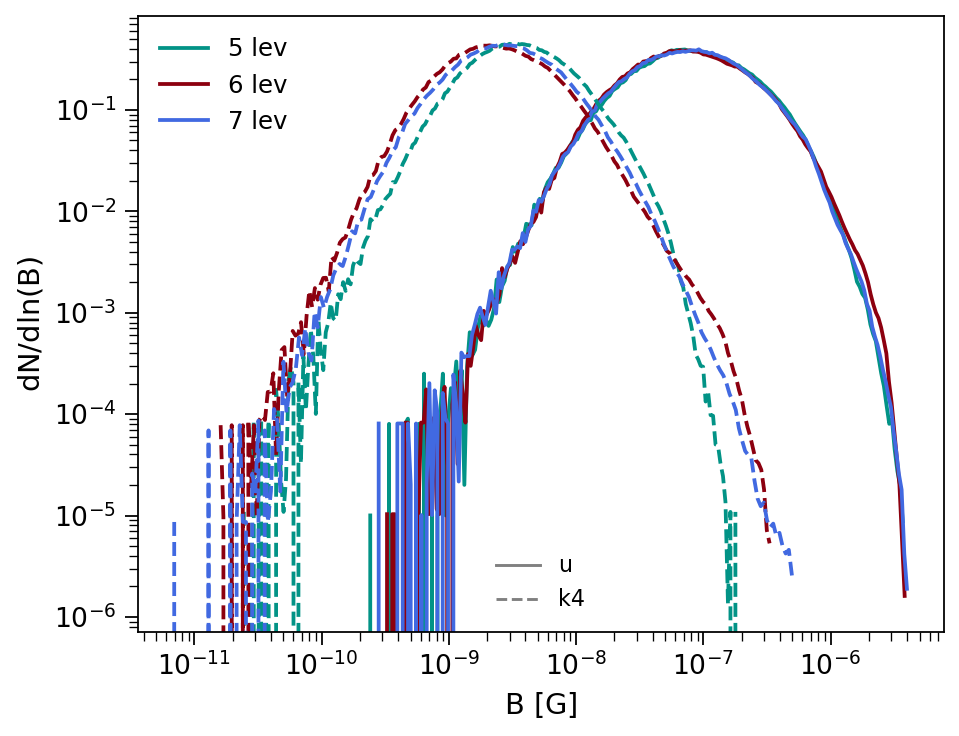}
    \caption{Upper panels: magnetic energy power spectra
    calculated for different AMR levels and different box sizes at $z=0$.
    We show the uniform (top panel) and Batchelor (middle panel) cases.
    The black dashed-dotted lines in each panel show the power
    spectra calculated from a zero-padded array.
    Lower panel: magnetic field PDFs of the uniform
    (solid lines) and Batchelor (dashed lines) models
    at $z=0$, at different AMR levels.
    } 
    \label{fig:Bps_resBoxPad}
\end{figure}

In this appendix, we discuss the dependence of our results on the adopted
spatial resolution.
We use the same initial conditions and perform different simulations, increasing the levels of AMR.
We show the results corresponding to a maximum of $5$, $6$ and $7$ levels
of AMR in Figure~\ref{fig:Bps_resBoxPad}.
The dependence on spatial resolution of the magnetic power spectra and
the PDFs of the magnetic field are shown by the different colors.
Even though we see greater variation for the Batchelor model
(the middle panel and the dashed lines of the bottom panel), we already observe
the convergence of both the uniform and Batchelor models already at the
sixth level of AMR and we see no significant changes in the shapes of the
magnetic energy spectra.

Spectral analysis based on Fourier transforms is a common approach
studying the scale dependence of the magnetic energy.
Nevertheless, some caveats to this approach result from the effects of
a limited box size and the nonperiodicity of the data.
In Figure~\ref{fig:Bps_resBoxPad}, we show the outcomes of these effects
on the magnetic energy spectra for the simulated uniform and Batchelor
models.
First, we see that for $k\la50\cMpchI$, corresponding to scales
$\gtrsim 20 \ckpch$, the spectra are well converged in the uniform model.
The shape of the magnetic spectra for both the uniform and Batchelor models 
are also 
mostly consistent with the spectra calculated in smaller/larger boxes. 
As expected, the amplitudes of the spectra are more strongly affected by
the size of the analyzed regions.
In particular, we see a $\sim$ one order of magnitude variation 
on the scales of $\sim 140 \ckpch$ for the uniform as well as Batchelor models. 

We also note that the nonperiodic boundary conditions of the selected box may distort the spectrum. In order to check this, we calculate the power spectra from the zero-padded array, extracted for the $(3 \cMpch)^3$ volume from the
five-level-AMR simulation (see the black dashed-dotted lines in
Figure~\ref{fig:Bps_resBoxPad}).
As seen in the figure, the power spectra calculated using the standard
method and zero padding lead to similar results, revealing that
our results as presented in the main text are not significantly affected
by the nonperiodicity of the data.

\begin{figure}[t]
    \centering
    \includegraphics[width=8.2cm]{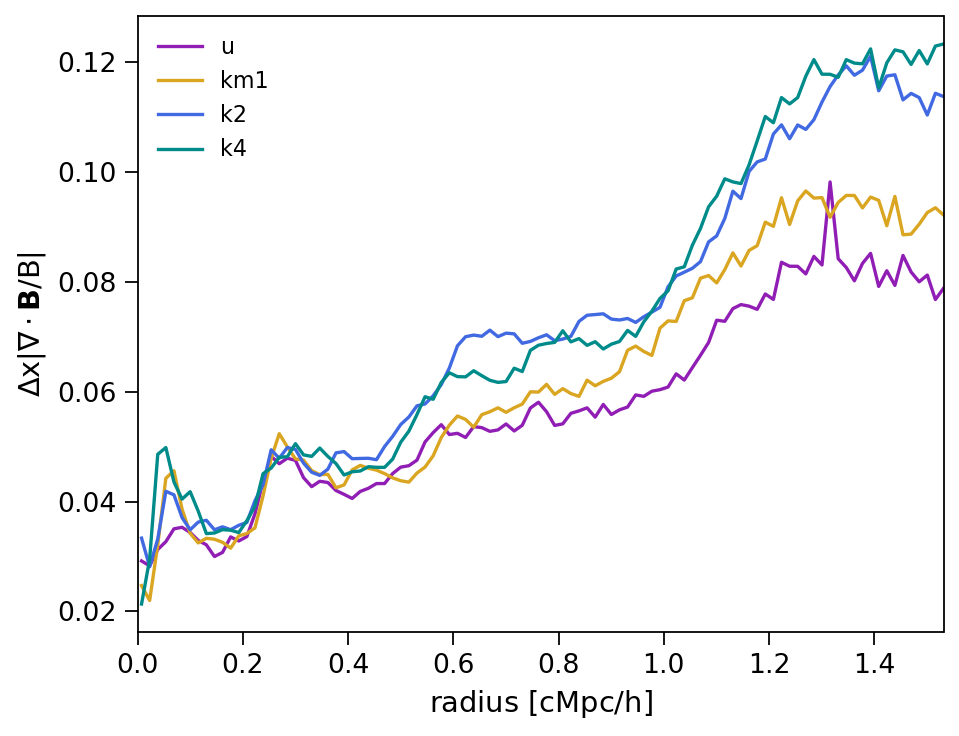}
    \caption{Normalized 
    divergences of the magnetic fields from the simulation with a maximum of seven levels of AMR (where $\Delta$ is the mesh spacing in the $x$-direction).
    } 
    \label{fig:divergence_radial}
\end{figure}

Finally, in Figure~\ref{fig:divergence_radial} we show the radial
profiles of the magnetic field divergence in our simulated cluster.
The largest differences between the models arise at $r \gtrsim 1\cMpch$, with the stochastic models having the largest values of divergence.
Nevertheless, as mentioned in Section~\ref{sec:NumAsp}, $\nabla\cdot\BB$
stays reasonably low in our four models in the largest fraction of the
simulated cluster volume. Quantitatively, we find that the normalized divergence remains below 10\%.

%

\section{Distribution of AMR levels}
\label{App:AMRlevelsProf}
%
\begin{figure}[htbp]
    \centering
    \includegraphics[width=8.2cm]{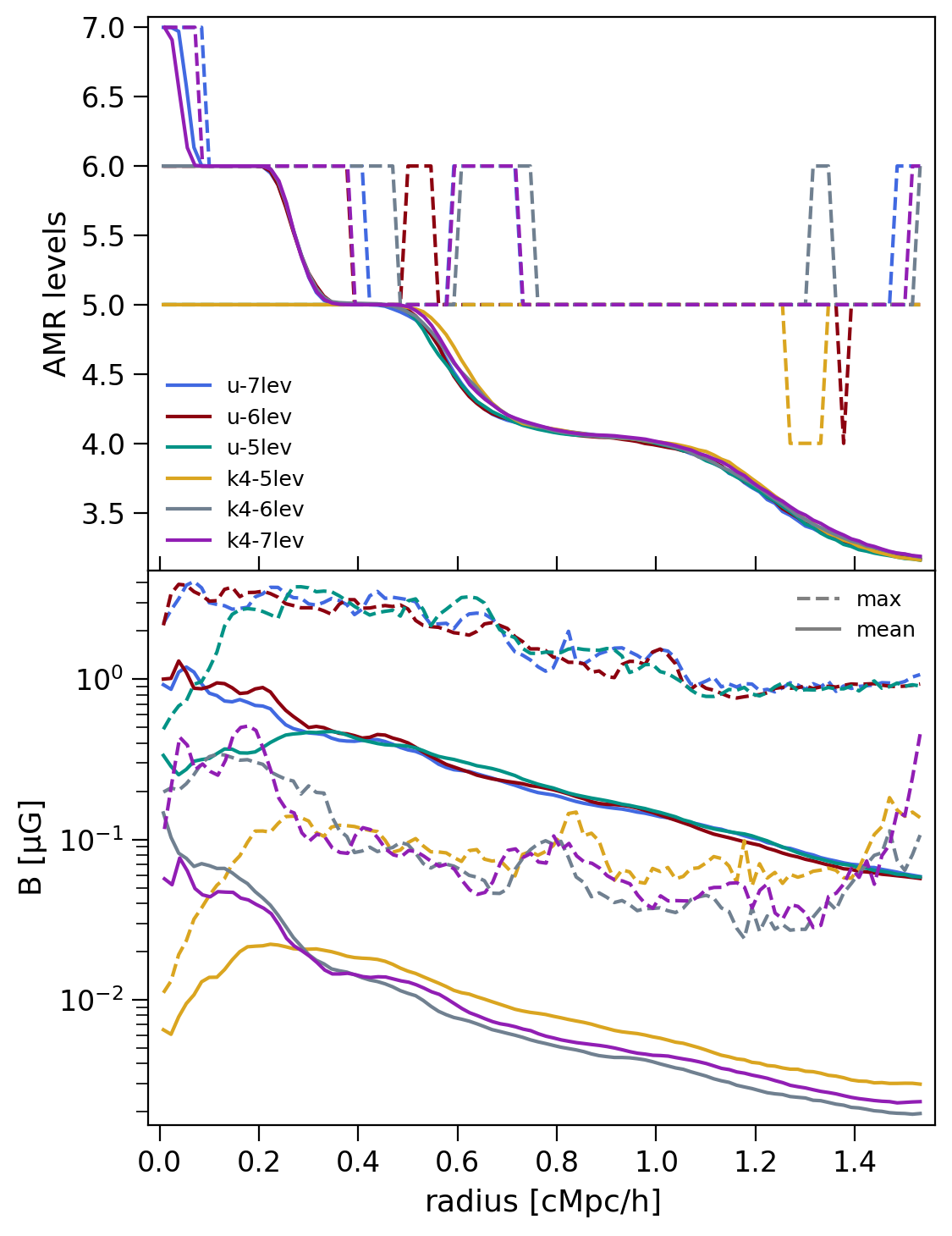}
    \caption{Radial distributions of the refinement levels and magnetic fields.  The mean and maximum within each radial bin are shown by the solid and dashed lines, respectively. The profiles are shown for the uniform and Batchelor models, calculated from a sphere with an $R_{\text{rvir}}$ radius.
     } 
    \label{fig:levels_radial1}
\end{figure}

Similar to \cite{Vazzaetal2018}, we show the radial profiles of the AMR levels along with the magnetic field 
profile in Figure~\ref{fig:levels_radial1}, for the uniform and Batchelor cases. In the top panel of Figure~\ref{fig:levels_radial1} we see 
that our simulated cluster is resolved
with a maximum of five AMR levels
(with $9.77 \ckpch$ resolution) 
in the $(1.5 \, \cMpch)^3$ central region, while the mean AMR level decreases 
toward the outskirts.  
On the other hand, the magnetic field profiles (the bottom panel of
Figure~\ref{fig:levels_radial1}) show larger strengths only in the cluster
core when the maximum levels of AMR are increased from five to seven.
Our AMR scheme is different from the one used in 
\cite{Vazzaetal2018} where the cluster is refined up to at least a sixth AMR level, even on the cluster outskirts.
An important difference, however, between the simulated clusters used
in this work and those used in \cite{Vazzaetal2018} is the mass of the cluster, which is one order of magnitude larger in the latter work.

In addition, we check the convergence of our AMR scheme by 
running an extra simulation with
a maximum of eight levels of AMR (for the Batchelor model, not shown). 
We do not see an important improvement in the AMR coverage of the cluster region when using higher levels of AMR. 
Therefore, given our selected refinement parameters, our AMR scheme already converges at six AMR levels.

\listofchanges
\end{document}